\pdfoutput=1

\documentclass[reprint,aps,pre,showpacstwocolumn]{revtex4-1}
\usepackage{mathptmx}
\usepackage{epsfig,amsopn}
\usepackage{graphicx}
\usepackage{amsmath,amssymb}
\usepackage{natbib}
\usepackage{xcolor}
\usepackage{braket}
\usepackage{float}
\usepackage{enumerate}
\usepackage{mathtools}
\usepackage[symbol]{footmisc}
\usepackage{appendix}
\usepackage{stackengine}
\usepackage{hyperref}
\allowdisplaybreaks[1]

\hypersetup{
    colorlinks=true,
    linkcolor=blue,
    citecolor=blue, 
    urlcolor=blue,
    }

\newcommand{\eref}[1]{Eq.~(\ref{#1})}%
\newcommand{\fref}[1]{Fig.~\ref{#1}} %

\def\bea{\begin{eqnarray}}
\def\eea{\end{eqnarray}}

\def\p{\partial}

\begin{document}

\title{Expediting Feller process with stochastic resetting}

\author{{\normalsize{}Somrita Ray}{\normalsize{}}}
\email{somrita@iittp.ac.in}

\affiliation{\noindent \textit{
Department of Chemistry, Indian Institute of Technology Tirupati, Tirupati 517619, India}}

\date{\today}
\begin{abstract}
\noindent
We explore the effect of stochastic resetting on the first-passage properties of Feller process. The Feller process can be envisioned as space-dependent diffusion, with diffusion coefficient $D(x)=x$, in a potential $U(x)=x\left(\frac{x}{2}-\theta \right)$ that owns a minimum at $\theta$. This restricts the process to the positive side of the origin and therefore, Feller diffusion can successfully model a vast array of phenomena in biological and social sciences, where realization of negative values is forbidden. In our analytically tractable model system, a particle that undergoes Feller diffusion is subject to Poissonian resetting, i.e., taken back to its initial position at a constant rate $r$, after random time epochs. We addressed the two distinct cases that arise when the relative position of the absorbing boundary ($x_a$) with respect to the initial position of the particle ($x_0$) differ, i.e., for (a) $x_0<x_a$ and (b) $x_a<x_0$. Utilizing the Fokker-Planck description of the system, we obtained closed-form expressions for the Laplace transform of the survival probability and hence derived the exact expressions of the mean first-passage time $\left<T_{r}\right>$. Performing a comprehensive analysis on the optimal resetting rate ($r^{\star}$) that minimize $\left<T_{r}\right>$ and the maximal speedup that $r^{\star}$ renders, we identify the phase space where Poissonian resetting facilitates first-passage for Feller diffusion. We observe that for $x_0<x_a$, resetting accelerates first-passage when $\theta<\theta_c$, where $\theta_c$ is a critical value of $\theta$ that decreases when $x_a$ is moved away from the origin. In stark contrast, for $x_a<x_0$, resetting accelerates first-passage when $\theta>\theta_c$, where $\theta_c$ is a critical value of $\theta$ that increases when $x_0$ is moved away from the origin. Our study opens up the possibility of a series of subsequent works with more case-specific models of Feller diffusion with resetting.
\end{abstract}




\maketitle
\section{Introduction}
The Feller process is a special kind of Markovian random process with a linear drift term and a state-dependent diffusion term, which vanishes at the origin \cite{FP1,FP2,FP3,FP4}. Such specific choices of the drift and diffusion terms ensure that the process is always restricted to the positive side of the origin. In other words, realization of negative values is absolutely forbidden for Feller diffusion, which in turn makes it a suitable model for describing a number of phenomena that are relevant in biological and social sciences. For example, the Feller process is frequently used as an alternative to the well-known Lotka-Volterra model \cite{LVM1,LVM2} to describe the time evolution of the population of a species in a locality, since it (Feller diffusion) includes the effect of fluctuating environment \cite{PD1,PD2,PD3}. One focal point of interest in these problems is to investigate the possibilities of extinction of that species and/or the unrestricted growth of its population, which can be extracted from the first-passage \cite{FPT1,persistence} properties of the Feller process. The Feller neuronal model \cite{Neuron1,Neuron2,Neuron3,Neuron4,Neuron5,Neuron6} is a simple yet effective one to recount the firing of single neurons. This integrate and fire model describes the fluctuations in membrane potential that regulates the nerve impulses; as this potential crosses a threshold value, the neuronal activity happens due to the firing of an action potential (nerve impulse), which lowers the membrane potential to some previous value and the cycle repeats. The firing dynamics can thus be explored by studying the associated first-passage properties of the Feller model \cite{Neuron1,Neuron2,Neuron3,Neuron4,Neuron5,Neuron6}.\\
\indent
Feller diffusion finds wide applications in the financial markets as well. For example, the well-celebrated CIR (Cox, Ingersoll and Ross) model \cite{FM1} is nothing but a Feller process that describes the temporal evolution of interest rate, where the randomness is originated solely from the market risk factor. Feller diffusion is also utilized for incorporating randomness in the volatility of asset prices \cite{FM2,FM3,FM4}, the latter being a statistical measure of the dispersion of the returns from those assets \cite{FM5}. For these reasons and others, Feller process has received a steady attention for the last few decades in biophysics and economics. \\
\indent
While Feller diffusion serves as a classic model to the cases mentioned above, there are quite a few situations where the original problem of Feller diffusion is not sufficient to explain the dynamics. For example, epidemics and natural disasters can abruptly diminish the population of a species in a geographic location, thereby setting it back to an earlier value \cite{population1}. In a similar way, during financial market crashes, the stock prices may drastically reduce to a prior asset value \cite{economics}. In these cases, Feller diffusion with {\it stochastic resetting} should serve as an excellent model. \\
\indent
Stochastic resetting \cite{SMReview,SM1,SM2,SM3} implies a situation where an ongoing dynamical process is stopped at random intervals of time, usually by some external protocol, to start over. Resetting can either shorten or prolong the completion of a stochastic process depending on the physical governing parameters. Tuning such parameters, it is (in principle) possible to invert such effect of resetting on the dynamics \cite{Restart-Biophysics1,ReuveniPRL,RayReuveniJPhysA,exponent,Landau,RRJCP,RayJCP,RRJCP-CM}. Due to its appearance in a plethora of natural systems and drastic effect on the dynamics, study of first-passage problems with resetting has gained overwhelming attention in recent years \cite{FPUR1,FPUR2,FPUR3,FPUR4,FPUR5,FPUR6,FPUR7,FPUR8,FPUR9,durang,FPUR10,FPUR11,FPUR12,FPUR13,SM4,FPUR14,FPUR15,FPUR16}. Surprisingly, the effect of resetting on Feller process still remains scarcely explored. To bridge this gap, in this paper we present a comprehensive analysis on the first-passage properties of Feller diffusion with resetting. \\
\indent
The rest of this paper is organized as follows. In Sec. II we discuss the equation of motion for the Feller process to visualize it as diffusion in a potential and revisit some earlier results associated to the first-passage properties of Feller diffusion without resetting. In the same Section, with the aid of the theory of first-passage with resetting \cite{Restart-Biophysics1,ReuveniPRL}, we predict when resetting is expected to accelerate first-passage for Feller diffusion. In Sec. III, we start with the Fokker Planck description of Feller diffusion with Poissonian resetting and obtain a general expression of the survival probability in the Laplace space, which depends on the boundary conditions. Considering that the target value is higher than the initial value of the associated first-passage process, in Sec. IV we first derive an exact, closed-form expression of the survival probability in the Laplace space, and then explore the phase space where resetting expedites Feller process. In Sec. V we repeat the entire study for the case where the target value is lower than the initial value of the Feller process. The final conclusions are drawn in Sec. VI.
\section{The Feller Process}
The Feller process \cite{FP1,FP2,FP3,FP4} is a special kind of Markovian random process with a linear deterministic force term (drift term) and a multiplicative noise term (diffusion term). Letting $X(t)$ denote a Feller process, its equation of motion can be written as
\begin{equation}
  d{X}(t)=a[b-X(t)]dt+\sigma\sqrt{X(t)}dW(t),
    \label{FP}
\end{equation}
where $a,b,\sigma>0$ are constant parameters and $W(t)$ is a Wiener process \cite{risken}, which represents the integral of a Gaussian white noise. \eref{FP} shows that the process has a linear drift term, $f(X,t)\coloneqq a[b-X(t)]$, and a space-dependent diffusion term $D(X,t)\coloneqq \sigma^2 X(t)$. Therefore, for $X(t)=0$, the drift term becomes $f(0,t)=ab>0$ (i.e., at the origin, the drift is directed towards its positive side) and the diffusion term becomes $D(0,t)=0$. These indicate that the Feller process $X(t)$ is always restricted to the positive side of the origin. \\
\indent
To simplify \eref{FP} further, we scale time as $t^{\prime}\to a t$ and the process as $x\to (2a/\sigma^2)X$ to rewrite \eref{FP} as
\begin{align}
  \frac{d{x}(t^{\prime})}{dt^{\prime}}=\theta-x(t^{\prime})+\sqrt{2x(t^{\prime})}\eta(t^{\prime}),
    \label{FP_scaled}
\end{align}
where $\theta\equiv 2ab/\sigma^2>0$ is the sole, constant and dimensionless governing parameter for the scaled process, which gives the rate of change of $x(t^{\prime})$ at the origin, i.e, $\theta=\left[dx(t^{\prime})/dt^{\prime}\right]_{x\to0}$. In \eref{FP_scaled}, $\eta(t^{\prime})\coloneqq dW(t^{\prime})/dt^{\prime}$ denotes a Gaussian white noise given by $\left<\eta(t^{\prime})\right>=0$ and $\left<\eta(t^{\prime})\eta(t^{\prime\prime})\right>=\delta(t^{\prime}-t^{\prime\prime})$.
This scaled description of the Feller diffusion will be considered throughout this paper. To avoid unnecessary complexity in the notation, we shall drop the prime and simply use $t$ to denote the scaled time variable from now on. Next, we briefly discuss how Feller process can be realized as inhomogeneous diffusion in a non-linear/non-monotonic potential.
\begin{figure}[ht!]
\begin{centering}
\includegraphics[width=8.2cm]{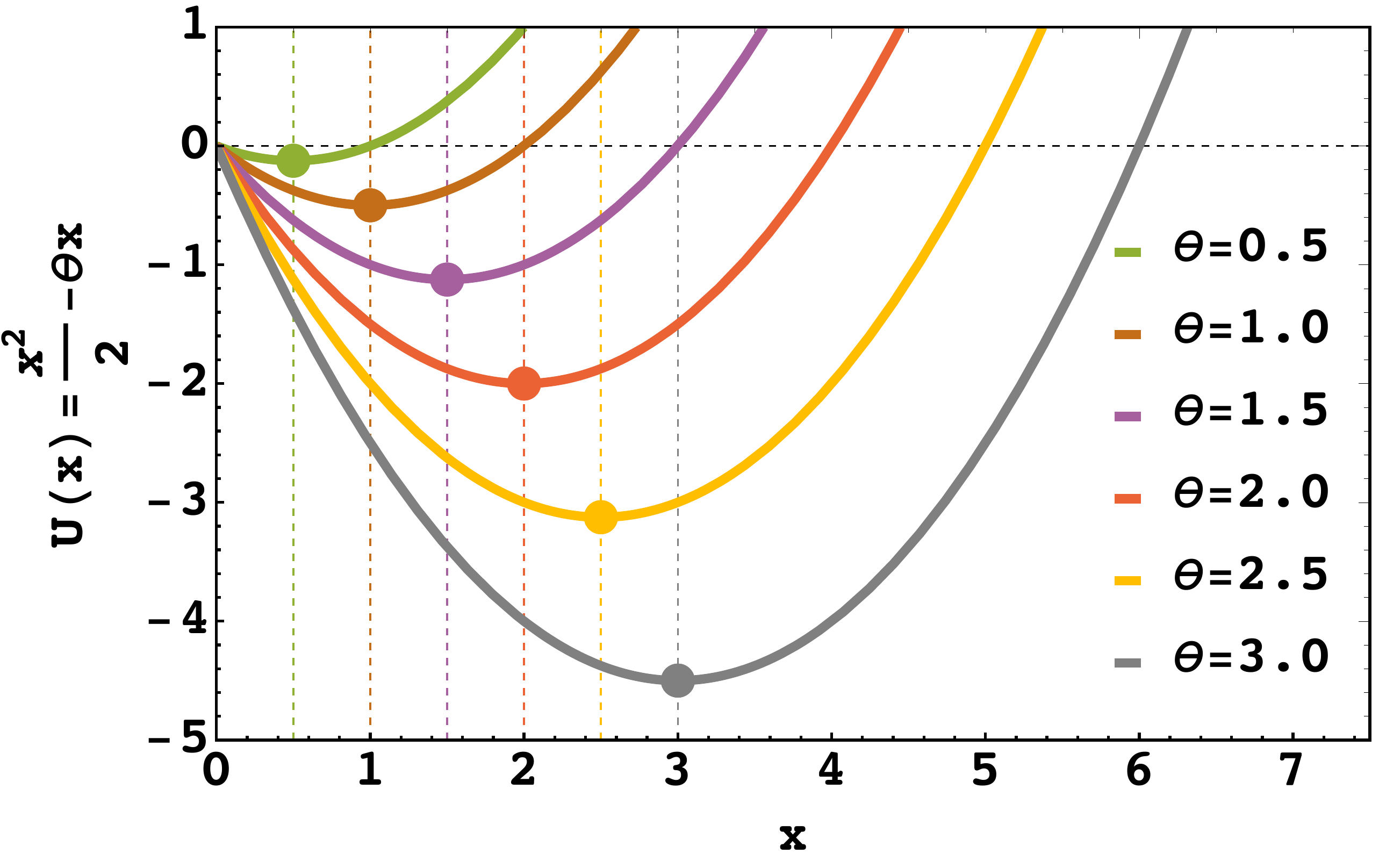}
\end{centering}
\caption{Feller process, as described in \eref{FP_scaled}, can be envisioned as diffusion in a potential $U(x)=x\left(\frac{x}{2}-\theta\right)$. As $\theta$ increases, the potential gradually becomes deeper and its minimum, which lies at $\theta$, shifts toward right. }
\label{Fig1}
\end{figure}
\subsection{Feller process as diffusion in a potential}
\eref{FP_scaled} clearly indicates that the scaled process is analogous to space-dependent diffusion of a Brownian particle with diffusion coefficient $D(x)=x$ in a force-field $-U^{\prime}(x)=(\theta-x)$ that vanishes at $x=\theta$ [also evident from \eref{fpe}], where $x$ is the dimensionless position of that diffusing particle. This force-field is generated from a potential $U(x)=x\left(\frac{x}{2}-\theta\right)$. Taking a more engaging look at the shape of $U(x)$, we see that it is a unique harmonic potential, where the equilibrium or minimum position lies at $\theta$. With increase in $\theta$, the potential becomes deeper and its minimum moves towards right [see \fref{Fig1}]. Note that the inhomogeneous diffusion invokes some  asymmetry in this otherwise symmetric potential $U(x)$. Since the diffusion at the origin is zero and it increases linearly with $x$, the effective potential that the particle experiences at the left side of $x=\theta$, the stable point of the potential, is always stronger compared to that at the right side of it. These features of Feller diffusion make the problem a rather complicated one to analyse, and we expect non-trivial outcomes in the first-passage properties. Indeed, depending on the relative placements of the initial position of the particle (denoted $x_0$) and the placement of the absorbing boundary (until which the first-passage is considered, denoted $x_a$) with respect to $\theta$, the Feller diffusion of interest can be either uphill or downhill (or a combination of both). Therefore, the relative placements of $\theta$, $x_0$ and $x_a$ will dictate whether the interplay between the drift velocity, generated from the potential, and the inhomogeneous diffusion will assist or oppose the first-passage to $x_a$. We will discuss this aspect in greater details later, while analysing the conditions where resetting facilitates first-passage. Now, we focus on constructing the Fokker-Planck equation associated to Feller diffusion in order to extract the first-passage properties. 
\begin{figure*}[ht!]
\begin{centering}
\includegraphics[width=7.6cm]{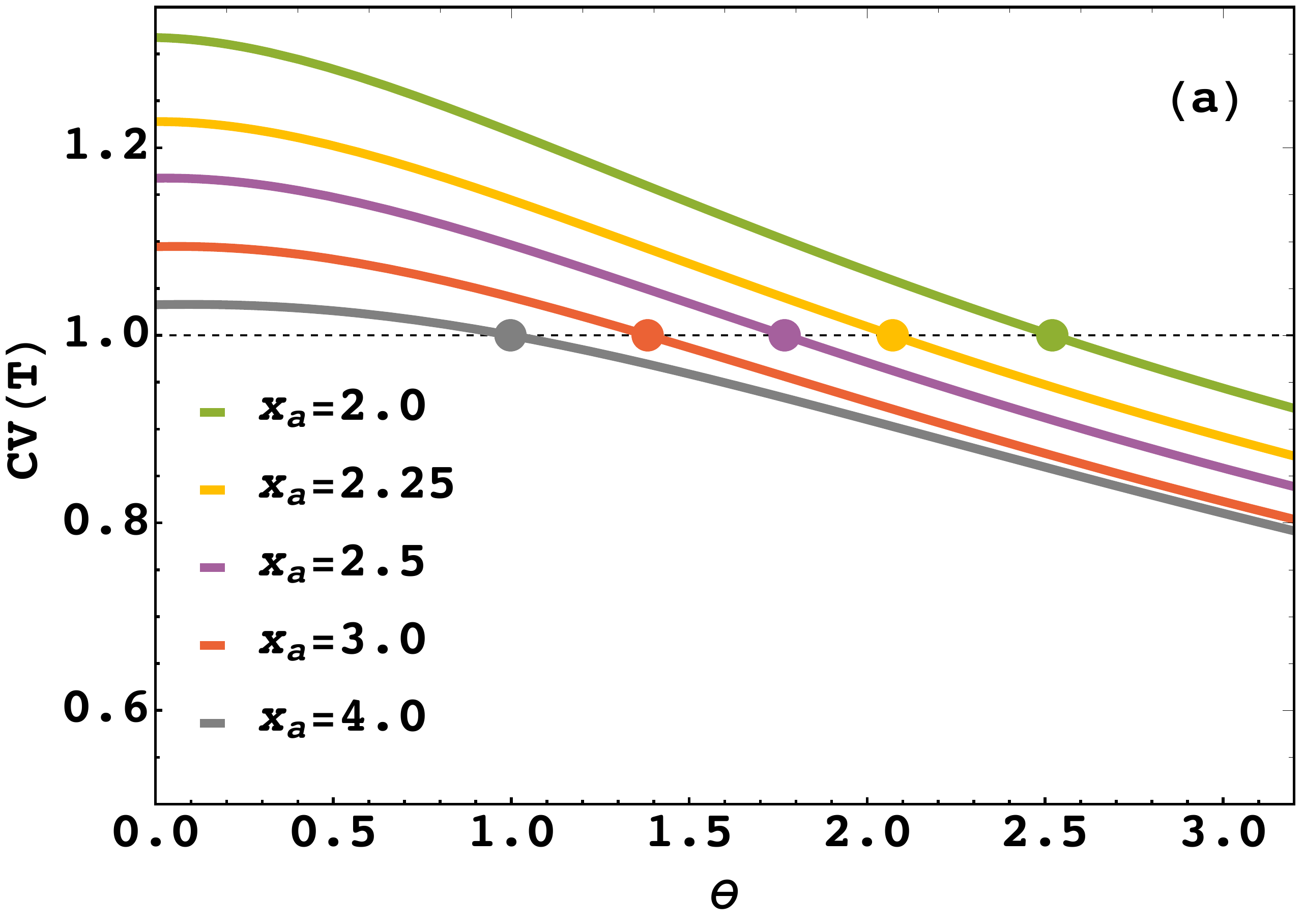}
\hspace{1.5cm}
\includegraphics[width=7.6cm]{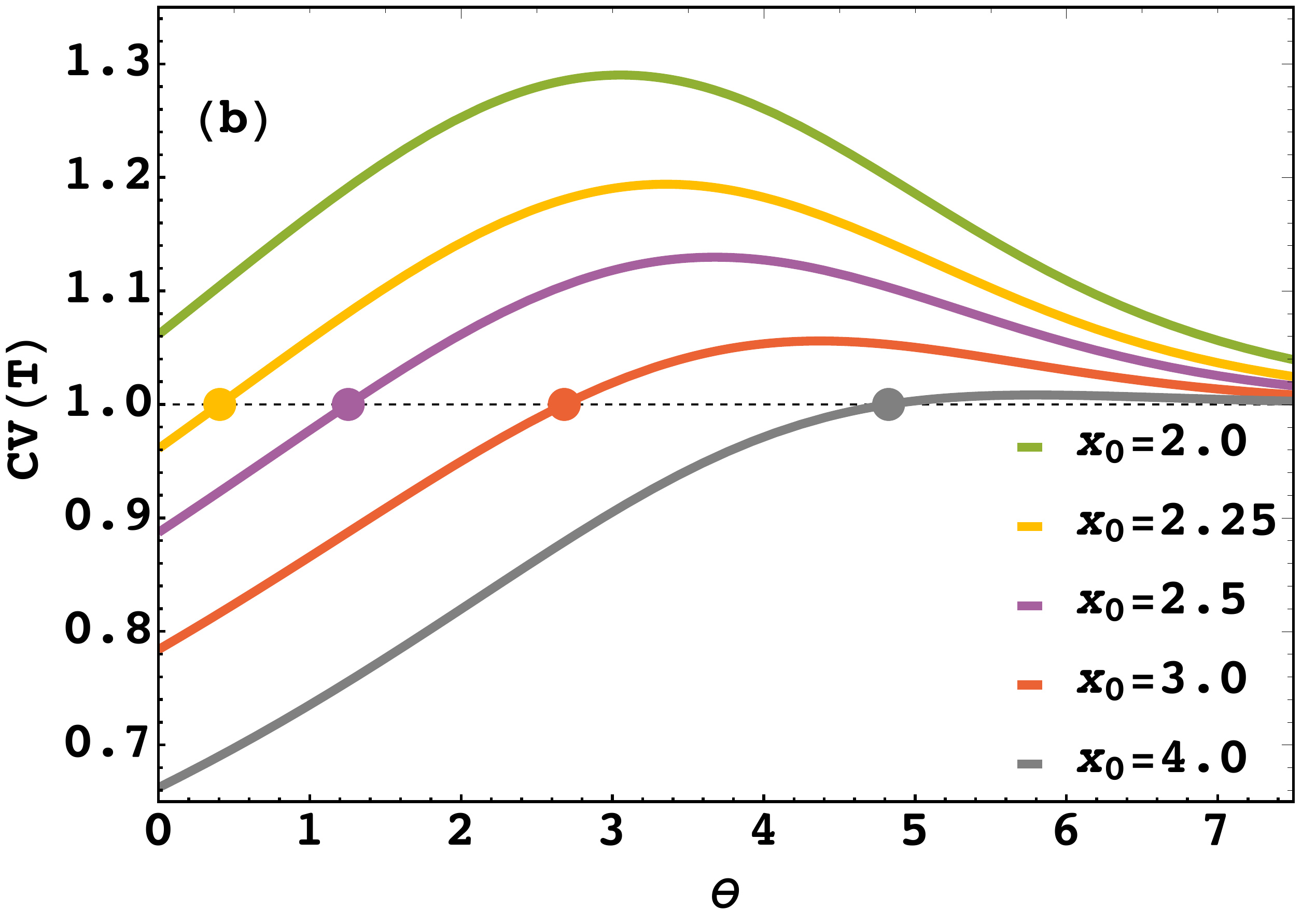}
\end{centering}
\caption{Panel(a): The coefficient of variation $CV(T)$ in the FPT vs. $\theta$ [following \eref{CVR} from Appendix A] for different values of $x_a$, keeping $x_0$ constant, where $x_0<x_a$ in all cases. $CV(T)$ is greater than unity for lower values of $\theta$, indicating that resetting should expedite first-passage from the initial position $x_0$ to the absorbing boundary at $x_a$ in that regime. Panel(b): The coefficient of variation $CV(T)$ in the FPT vs. $\theta$ [following \eref{CVL} from Appendix A] for different values of $x_0$, keeping $x_a$ constant, where $x_a<x_0$ in all cases. In contrast to the previous case, $CV(T)$ is greater than unity for higher values of $\theta$, which indicates that resetting should expedite first-passage from $x_0$ to $x_a$ in that regime.
}
\label{Fig2}
\end{figure*}
\subsection{Fokker-Planck equation and  survival probability}
\indent
The Fokker-Planck equation \cite{risken,SMReview} associated to \eref{FP_scaled} reads
\begin{align}
  \frac{\p }{\p t}p(x,t|x_0)=\frac{\p }{\p x}[(x-\theta)p(x,t|x_0)]+\frac{\p^2 }{\p x^2}[x p(x,t|x_0)],
    \label{fpe}
\end{align}
where $p(x,t|x_0)$ is the conditional probability density function for the process to be at position $x$ at time $t$, the initial condition being $p(x,0|x_0)=\delta(x-x_0)$. Since $0<x<\infty$ represent the natural boundaries for Feller diffusion, the initial condition uniquely specifies $p$. Note that to construct \eref{fpe} from \eref{FP_scaled}, we follow the It\^{o} convention  \cite{vankampen}. From \eref{fpe}, we can write the backward Fokker-Planck equation \cite{gardinar,SMReview} in terms of the survival probability, i.e., the total probability density that the process survives within an interval $\Omega$ at time $t$, $Q(t|x_0)\coloneqq\int_{\Omega}p(x,t|x_0)dx$, which reads
\begin{align}
    \frac{\p }{\p t}Q(t|x_0)=(\theta-x_0)\frac{\p }{\p x_0}Q(t|x_0)+x_0\frac{\p^2 }{\p x_0^2}Q(t|x_0).
    \label{bfpe_Q}
    \end{align}
Recently, the first-passage properties of a Feller process have been explored \cite{FP4}, where it was shown that the first-passage to a threshold value $x_a$ that is above the initial value $x_0$, the survival probability in the Laplace space is given by
 \begin{align}
    \tilde{Q}(s|x_0)=
    \frac{1}{s}\left[1-\frac{M(s;\theta;x_0)}{M(s;\theta;x_a)}\right],
    \label{Q_solR}
\end{align}  
where $\tilde{Q}(s|x_0)\coloneqq\int_{0}^{\infty}e^{-st}Q(t|x_0)dt$ denotes the Laplace transform of $Q(t|x_0)$, $s$ being the Laplace variable. $M(s;\theta;y)$ in \eref{M_def} is the confluent hypergeometric function of the first kind \cite{DLMF}, defined as 
\begin{align}
    M(s;\theta;y)\coloneqq\sum_k^{\infty}\frac{\Gamma(s+k)}{\Gamma(\theta+k)}\frac{y^k}{k!},
    \label{M_def}
\end{align}
where $\Gamma(c)\coloneqq\int_0^{\infty}t^{c-1}e^{-t}dt$ is the \textit{gamma function}. It has also been established that when the threshold value $x_a$ lies below the initial value $x_0$, the survival probability in the Laplace space is given by 
\begin{align}
    \tilde{Q}(s|x_0)=
    \frac{1}{s}\left[1-\frac{U(s;\theta;x_0)}{U(s;\theta;x_a)}\right]
    \label{Q_solL},
\end{align}
where $U(s;\theta;y)$ is the confluent hypergeometric function of the second kind \cite{DLMF}, expressed in terms of $M(s;\theta;y)$ as
\begin{align}
    U(s;\theta;y)\coloneqq \pi\csc{(\pi \theta)}\left[
    -y^{1-\theta}\frac{M(s-\theta+1;2-\theta;y)}{\Gamma(s)}\right.\nonumber\\
   +\left.\frac{M(s;\theta;y)}{\Gamma(s-\theta+1)}
    \right].
    \label{U_def}
\end{align}
The survival probability contains complete information of the associated first-passage process. In particular, when a process takes a random time $T$ to complete, the first and second moments of that time can be calculated from its survival probability in the Laplace space \cite{gardinar} as $\left<T\right>=\left[\tilde{Q}(s|x_0)\right]_{s\to 0}$ and 
$\left<T^2\right>=-2\left[\frac{d\tilde{Q}(s|x_0)}{ds}\right]_{s\to0}$, respectively. Next, we briefly discuss how one can predict whether the introduction of stochastic resetting can reduce that mean time of completion or not, based on these two quantities.
\subsection{Can stochastic resetting accelerate Feller diffusion?}
Stochastic resetting, i.e., bringing a diffusing particle back to its initial position after random intervals of time, can either facilitate or hinder a first-passage process. The theory of first-passage with resetting \cite{Restart-Biophysics1,ReuveniPRL} states that resetting expedites a first-passage process whenever the standard deviation of the associated first-passage time (FPT), $\sigma(T)\coloneqq[\left<T^2\right>-\left<T\right>^2]^{\frac{1}{2}}$, is greater than the mean first-passage time, $\left<T\right>$. In complete contrast, when $\sigma(T)<\left<T\right>$, resetting delays first-passage. Since the coefficient of variation in FPT is defined as $CV(T)\coloneqq\sigma(T)/\left<T\right>$, one can alternatively say that whenever the $CV(T)>1$, resetting reduces the mean FPT of the process, otherwise (i.e., when $CV(T)<1$) the original process is hindered due to resetting and the associated mean FPT increases as a result. In physical systems, the mean FPT, $\left<T\right>$, and the fluctuations around it, quantified by $\sigma(T)$, both vary when the governing parameters are altered. This indicates that by tuning physical parameters, the effect of resetting on the dynamics can be inverted. \\
\indent
In order to get a qualitative idea about how resetting affects the first-passage for Feller diffusion, we calculate the associated $CV(T)$ for the following two conditions. First, utilizing \eref{Q_solR} we calculate $CV(T)$ for $x_0<x_a$ [see Appendix A for details], and plot the same in \fref{Fig2}(a) with respect to $\theta$ for different values of $x_a$, keeping $x_0$ constant. \fref{Fig2}(a) clearly shows that $CV(T)>1$ for smaller values of $\theta$, which indicates that resetting expedites first-passage there. When $\theta$ increases beyond a threshold value, however, $CV(T)$ decreases below unity, which means resetting delays first-passage in that regime. Next, utilizing \eref{Q_solL} we calculate $CV(T)$ as a function of $\theta$ for $x_a<x_0$, for different values of $x_0$ keeping $x_a$ constant [see Appendix A for details]. Plotting the same in \fref{Fig2}(b), we observe that in stark contrast to the previous case, here resetting hinders first-passage for smaller values of $\theta$ and accelerates the same as $\theta$ grows beyond a critical value. These observations suggest that based on the governing parameters, viz., $\theta$, $x_a$ and $x_0$, resetting can either expedite or delay first-passage for Feller process. While this condition based on $CV(T)$ gives a general idea about the regimes where resetting facilitates (or hinders) first-passage, it fails to comment on the quantitative aspect of the resulting speedup (or delay). Nonetheless, it suggests that Feller diffusion can lead to non-trivial first-passage properties when subject to resetting. Motivated by these initial findings, we now perform a comprehensive analysis on the effect of Poissonian resetting on the dynamics of Feller diffusion. 
\section{Feller Process with resetting}
Consider a particle executing Feller diffusion in one dimension following \eref{FP_scaled}, starting from a position $x_0>0$. In addition, assume that by some external protocol, it is being stochastically reset to a position $x_r>0$ at a constant rate $r$. This means that the random times between two subsequent resetting events are chosen from an exponential distribution with mean $r^{-1}$. Consider an absorbing boundary placed at $x_a>0$; when the particle hits it for the first time, it gets absorbed and the process is complete. Letting $p_r(x,t|x_0)$ denote the conditional probability density of finding the particle at position $x$ at time $t$, provided its initial position was $x_0$, we write down the forward Fokker Planck equation \cite{SMReview,RayReuveniJPhysA,RRJCP,RayJCP} for the process as
\begin{align}
  \frac{\p }{\p t}p_r(x,t|x_0)=\frac{\p }{\p x}[(x-\theta)p_r(x,t|x_0)]+\frac{\p^2 }{\p x^2}[x p_r(x,t|x_0)]\nonumber\\
    -rp_r(x,t|x_0)+r\delta(x-x_r)Q_r(t|x_0),
    \label{FP_ffpe}
\end{align}
where $Q_r(t|x_0)\coloneqq\int_{\Omega}p_r(x,t|x_0)dx$ is the survival probability within the interval $\Omega$ (i.e., the total probability of finding the particle within $\Omega$ at time $t$). Note that the placement of the absorbing boundary with respect to the initial position dictates the interval that the particle is within. For $x_0<x_a$, $\Omega=[0,x_a]$ and for $x_a<x_0$, $\Omega=[x_a,\infty)$.\\
\indent
It is evident from \eref{FP_ffpe} that in the absence of resetting, it boils down to \eref{FP}, the Fokker Planck equation of the original Feller process. For $r>0$,  probability of being at position $x$ decreases and that at $x_r$ increases due to resetting, and the two additional terms appear in \eref{FP_ffpe} to account for this additional probability flow. \\
\indent
The backward Fokker Planck equation in terms of the survival probability $Q_r(t|x_0)$ thus reads \cite{SMReview,RRJCP,RayJCP} 
\begin{align}
    \frac{\p }{\p t}Q_r(t|x_0)=(\theta-x_0)\frac{\p }{\p x_0}Q_r(t|x_0)+x_0\frac{\p^2 }{\p x_0^2}Q_r(t|x_0)\nonumber\\
    -rQ_r(t|x_0)+rQ_r(t|x_r).
    \label{bfpe_Qr}
\end{align}
Laplace transforming \eref{bfpe_Qr} and utilizing the initial condition, $Q_r(0|x_0)=1$, we get
\begin{align}
    x_0\frac{\p^2 }{\p x_0^2}\tilde{Q}_r(s|x_0)+(\theta-x_0)\frac{\p }{\p x_0}\tilde{Q}_r(s|x_0)-(s+r)\tilde{Q}_r(s|x_0)\nonumber\\
    =-[1+r\tilde{Q}_r(s|x_r)],
    \label{bfpe_Qr_lt}
\end{align}
where $\tilde{Q}_r(s|x_0)\coloneqq\int_0^{\infty}e^{-st}Q_r(t|x_0)dt$ denotes the Laplace transform of $Q_r(t|x_0)$. To convert the non-homogeneous differential equation shown in \eref{bfpe_Qr_lt} to a homogeneous one, consider a constant shift 
\begin{align}
    \tilde{q}_r(s|x_0)=\tilde{Q}_r(s|x_0)-\frac{1+r\tilde{Q}_r(s|x_r)}{s+r}.
    \label{qr_lt}
\end{align}
This allows us to write down \eref{bfpe_Qr_lt} in terms of $\tilde{q}_r(s|x_0)$ as 
\begin{align}
    x_0\frac{\p^2 }{\p x_0^2}\tilde{q}_r(s|x_0)+(\theta-x_0)\frac{\p }{\p x_0}\tilde{q}_r(s|x_0)-(s+r)\tilde{q}_r(s|x_0)=0.
    \label{bfpe_qr}
\end{align}
\eref{bfpe_qr} resembles Kummer's equation  \cite{arfken,magnus}, which is a confluent hypergeometric equation with general solution
\begin{align}
    \tilde{q}_r(s|x_0)=c_1M(s+r;\theta;x_0)+c_2U(s+r;\theta;x_0).
    \label{qr_sol}
\end{align}
\begin{figure*}[ht!]
\begin{centering}
\includegraphics[width=7.5cm]{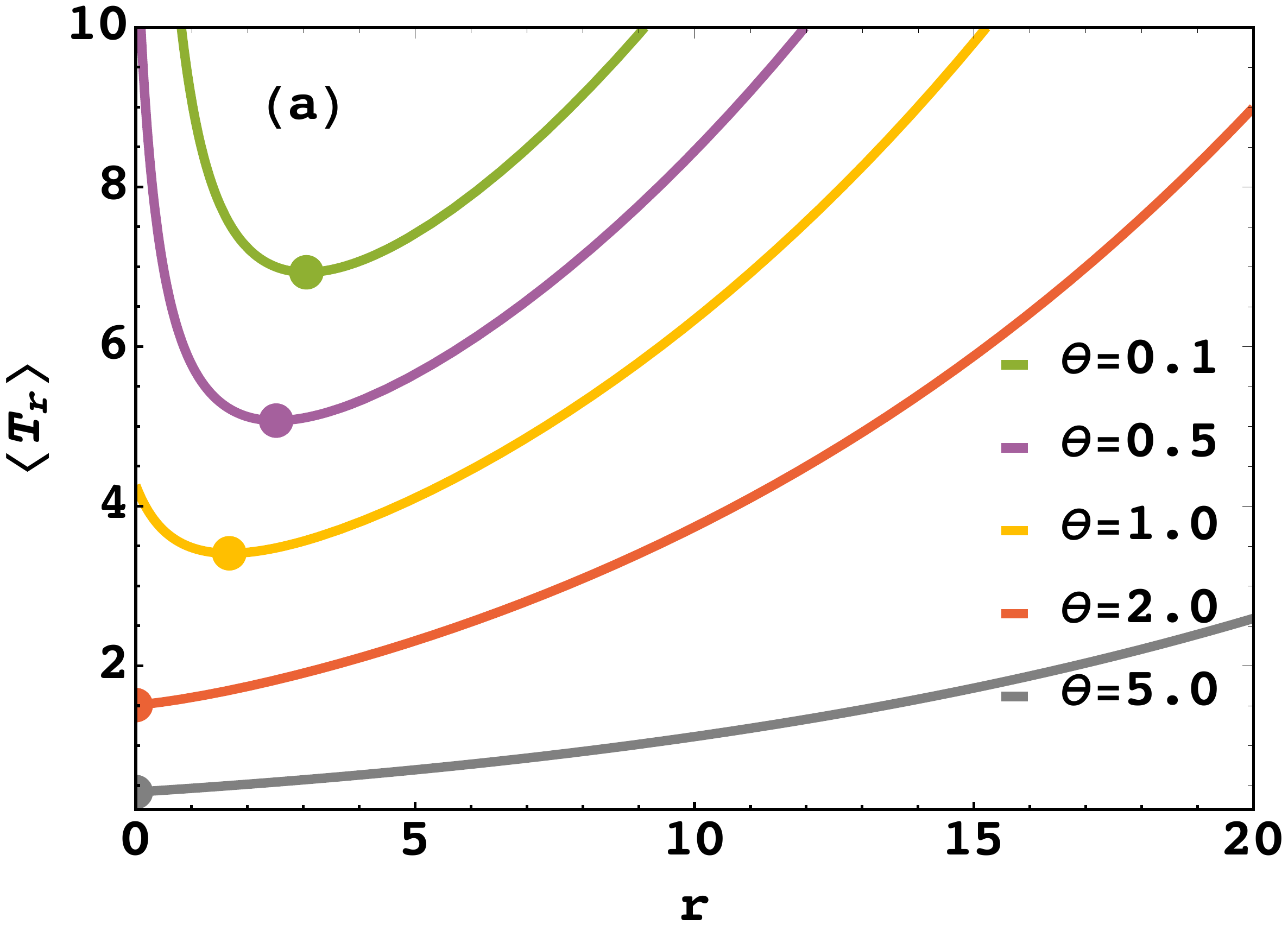}
\hspace{1.25cm}
\includegraphics[width=8.0cm]{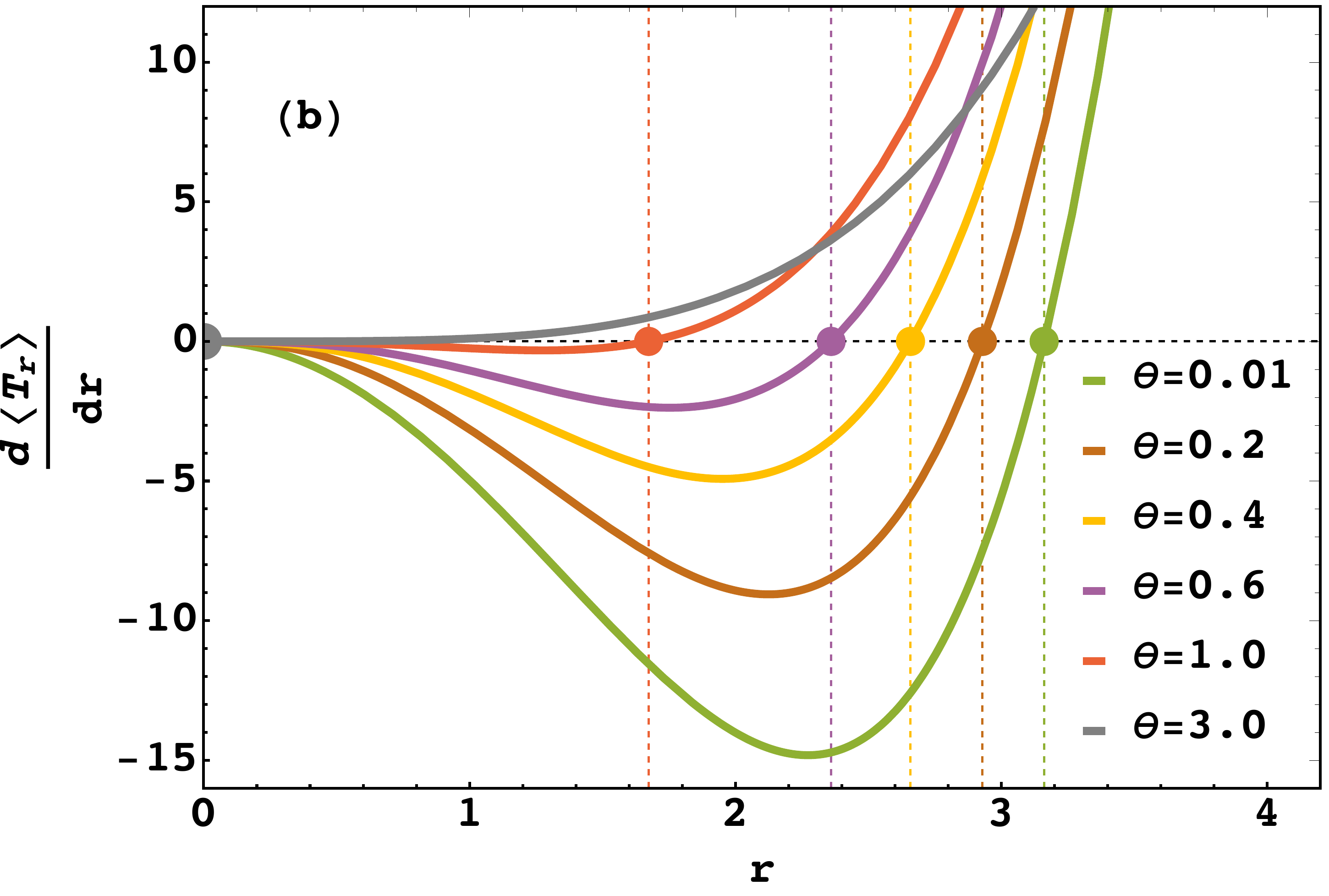}
\end{centering}
\caption{First-passage with resetting for $x_0<x_a$. Panel (a): The mean FPT $\left<T_r\right>$ vs. the resetting rate $r$ for different values of $\theta$, where $x_0=1$ and $x_a=2.5$. $\left<T_r\right>$ shows non-monotonic variation with $r$ for lower values of $\theta$; the optimal resetting rates ($r^{\star}$), marked by colored discs, are non-zero in this regime. In contrast, for higher values of $\theta$, the variation of $\left<T_r\right>$ with $r$ becomes monotonic and $r^{\star}$ become zero in this regime.
Panel (b): Graphical solution of \eref{Trans_R} for different values of $\theta$, where $x_0=1$ and $x_a=2.5$. The solutions that present the optimal resetting rate ($r^{\star}$) in each case, are marked by colored discs. 
}
\label{Fig3}
\end{figure*}
Here $M(s+r;\theta;x_0)$ and $U(s+r;\theta;x_0)$ are the confluent hypergeometric functions of the first and second kind \cite{DLMF}, as introduced in \eref{M_def} and \eref{U_def}, respectively. Combining \eref{qr_lt} and \eref{qr_sol} together, we can write down the general solution of \eref{bfpe_Qr_lt} that reads
\begin{eqnarray}
    \tilde{Q}_r(s|x_0)=c_1M(s+r;\theta;x_0)&+&c_2U(s+r;\theta;x_0)\nonumber\\
    &+&\frac{1+r\tilde{Q}_r(s|x_r)}{s+r}.
    \label{QrL_sol_g}
\end{eqnarray}
To find out the specific solution of \eref{bfpe_Qr_lt}, we need to calculate $c_1$ and $c_2$ from the boundary conditions. The absorbing boundary at $x_a$ leads to $\tilde{Q}_r(s|x_a)=0$. However, the specific solutions of \eref{bfpe_Qr_lt} will depend on the placement of $x_a$ with respect to the initial position $x_0$, as mentioned above. Once 
$\tilde{Q}_r(s|x_0)$ is calculated for the appropriate scenario, we can utilize that solution to calculate the first-passage time of the particle from $x_0$ to $x_a$, denoted $T_r$, in the following manner. Recall that the probability density of $T_r$ is given by $-dQ_r(t|x_0)/dt$, which allows us to calculate the moments of $T_r$ following the general relation \cite{gardinar} $\left<T_r^n\right>=-\int_0^{\infty}t^n\left[\frac{dQ_r(t|x_0)}{dt}\right]dt\equiv n(-1)^{n-1}\left[\frac{d^{n-1}\tilde{Q}_r(s|x_0)}{ds^{n-1}}\right]_{s=0}$. \\
\indent
In Section IV, we focus on the case $x_0<x_a$, i.e., where the boundary is placed at the right hand side of the initial position, and explore the first-passage from the initial position $x_0$ to the boundary at $x_a$.
\section{First-Passage from $x_0$ to $x_a$: when the absorbing boundary is placed further away from the origin compared to the initial position ($x_0<x_a$)}
Consider the case where $x_0<x_a$, i.e., when the particle diffuses in the interval $\Omega=[0,x_a]$. Note that the placement of the boundary further away from the origin compared to the initial position suggests that in this case, the first-passage is being considered from a less diffusive to a more diffusive zone. Going back to \eref{QrL_sol_g}, we see that in the limit $x_0\to 0$, $U(s+r;\theta;x_0)$ diverges for $\theta>1$. Hence we set $c_2=0$ to keep $\tilde{Q}_r(s|x_0)$ finite, irrespective of the values of $\theta$. The absorbing boundary at $x_a$ then leads to  $c_1=-\left[1+r\tilde{Q}_r(s|x_r)\right]/[(s+r)M(s+r;\theta;x_a)]$. The specific solution of \eref{bfpe_Qr_lt} for $x_0<x_a$ thus becomes
\begin{align}
    \tilde{Q}_r(s|x_0)=\frac{1+r\tilde{Q}_r(s|x_r)}{s+r}\left[1-\frac{M(s+r;\theta;x_0)}{M(s+r;\theta;x_a)}\right].
    \label{Qr_solR}
\end{align}
Setting $x_r=x_0$ in \eref{Qr_solR}, i.e, equating the position of reset to the initial position, allows us to obtain an explicit expression of $\tilde{Q}_r(s|x_0)$, which reads
\begin{align}
    \tilde{Q}_r(s|x_0)=
    \frac{1-\frac{M(s+r;\theta;x_0)}{M(s+r;\theta;x_a)}}{s+r \left[\frac{M(s+r;\theta;x_0)}{M(s+r;\theta;x_a)}\right]}.
    \label{Qr_sol2R}
\end{align}
Note that in the absence of resetting, i.e, when $r\to 0$, the survival probability in the Laplace space for $x_0<x_a$ reduces to the expression given in \eref{Q_solR}. \\
\indent
The mean first-passage time from $x_0$ to an absorbing boundary at $x_a$ can be obtained from \eref{Qr_sol2R} as $\left<T_r\right>=[\tilde{Q}_r(s|x_0)]_{s=0}$, which gives
\begin{equation}
    \left<T_r\right>=\frac{1}{r}\left[\frac{M(r;\theta;x_a)}{M(r;\theta;x_0)}-1\right].\\
    \label{MFPT_R}
\end{equation}
In \fref{Fig3}(a), we plot the mean FPT $\left<T_r\right>$ vs. the resetting rate $r$ following \eref{MFPT_R}, keeping $x_0$ and $x_a$ constant. \fref{Fig3}(a) shows that for lower values of $\theta$, $\left<T_r\right>$ varies non-monotonically with $r$; when $r$ is small, $\left<T_r\right>$ decreases as $r$ grows, but for higher values of the resetting rate, the mean FPT increases with $r$. A minimum in mean FPT for an {\it optimal} resetting rate is thus observed. As $\theta$ increases beyond a critical value [denoted $\theta_c$, not shown in \fref{Fig3}(a)],  however, the mean FPT monotonically increases with the resetting rate, which indicates that resetting can not expedite first-passage in that case. These two distinct types of variation of $\left<T_r\right>$ with $r$ show a hallmark of {\it resetting transition} \cite{RayReuveniJPhysA,exponent,Landau,RRJCP,RayJCP}. \fref{Fig3}(a) also shows that the optimal resetting rate, i.e, the resetting rate that corresponds to the minimum mean FPT, is zero when the variation of $\left<T_r\right>$ with $r$ is monotonic. When $\left<T_r\right>$ shows non-monotonic variation with $r$, the optimal resetting rate increases as $\theta$ becomes smaller. This suggests that the optimal resetting rate, denoted $r^{\star}$, should serve as an excellent observable in exploring the resetting transition for the present problem. Next, we calculate $r^{\star}$ as a function of $\theta$ to understand the resetting transition in greater depth.
\vspace{-0.2cm}
\subsection{The optimal resetting rate for $x_0<x_a$}
Since the optimal resetting rate minimizes the mean FPT, the rate of change of the mean FPT with the resetting rate becomes zero at $r=r^{\star}$, i.e,  $\left[\frac{d \left<T_r\right>}{dr}\right]_{r=r^{\star}}=0$. Therefore, differentiating \eref{MFPT_R} with respect to $r$ and equating that to zero for $r=r^{\star}$, we obtain 
\begin{widetext}
\begin{eqnarray}
\left[\frac{d \left<T_r\right>}{dr}\right]_{r=r^{\star}}\mbox{\hspace{-0.5cm}}=
\frac{1}{(r^{\star})^2 M(r^{\star};\theta;x_0)}
\left[M(r^{\star};\theta;x_0)\left(M(r^{\star};\theta;x_0)
-M(r^{\star};\theta;x_a)+r^{\star}\left[\frac{\p M(r;\theta;x_a)}{\p r}\right]_{r=r^{\star}}\right)\right.
\nonumber\\
\left.-r^{\star}M(r^{\star};\theta;x_a)
\left[\frac{\p M(r;\theta;x_0)}{\p r}\right]_{r=r^{\star}}\right]
=0.
    \label{Trans_R}
\end{eqnarray}
\end{widetext}
\eref{Trans_R} is a transcendental equation, hence not solvable analytically. \fref{Fig3}(b) shows its graphical solution, where the left hand side of \eref{Trans_R} is plotted against $r$ for different values of $\theta$, and the points of intersection of each curve with the abscissa
\begin{figure}[hb!]
\includegraphics[width=7.2cm]{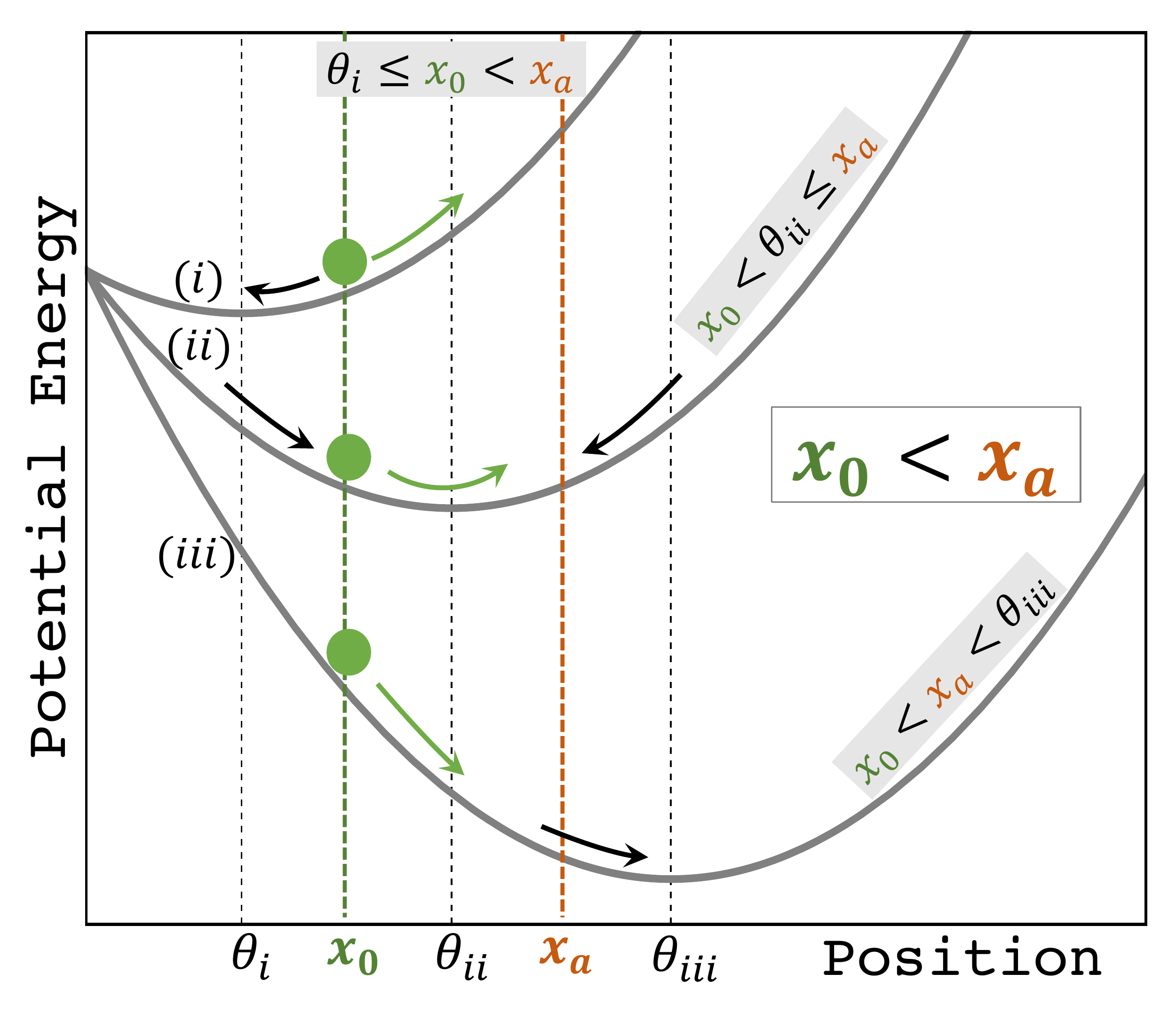}
\caption{Schematic plot of the potential $U(x)=x\left(\frac{x}{2}-\theta\right)$ vs. $x$ to envision Feller process as space-dependent diffusion in $U(x)$. The three scenarios that arise in connection to the first-passage from $x_0$ to $x_a$, where $x_0<x_a$, are illustrated with three $U(x)$ curves with different minima, such that $\theta_i<\theta_{ii}<\theta_{iii}$. The green arrow in each case indicates the overall direction of the first-passage, whereas the black arrows indicate the drift velocity generated by $U(x)$ in each case. When these two forces oppose each other [e.g., $\theta_i\le x_0<x_a$], resetting is expected to expedite first-passage. In contrast, when these forces assist each other [e.g., $x_0<x_a<\theta_{iii}$], resetting is expected to delay first-passage. An intermediate case is observed for $x_0<\theta_{ii}\le x_a$.
}
\label{Fig4}
\end{figure}
give the corresponding optimal resetting rate. As observed from \fref{Fig3}(b), the optimal resetting rates are higher for smaller values of $\theta$, indicating that resetting accelerates the first-passage when $\theta<\theta_c$. As $\theta$ grows, $r^{\star}$ decreases and finally becomes zero for $\theta\ge\theta_c$, which means that there resetting can not expedite the first-passage. This trend can be qualitatively understood by considering Feller process as space-dependent diffusion, with diffusion coefficient $D(x)=x$, in a potential $U(x)=x\left(\frac{x}{2}-\theta\right)$, as discussed in Section II. If we focus on the relative placements of $\theta$ with $x_0$ and $x_a$, we see that three distinct possibilities arise: (i) $\theta\le x_0<x_a$, (ii) $x_0<\theta\le x_a$ and (iii) $x_0<x_a<\theta$. In \fref{Fig4}, we plot the potential energy $U(x)$ vs. the position $x$ to illustrate these three cases. Recalling that the minimum of the potential $U(x)$ lies at $\theta$, the first-passage for case (i) is clearly an uphill journey [marked by the green arrow above curve (i)], whereas for case (iii) it is a downhill one [marked by the green arrow above curve (iii)]. In other words, for case (i) the drift velocity acts away from the absorbing boundary [marked by the black arrow above curve (i)] and thereby opposes the first-passage; resetting at $x_0>\theta$ thus helps accelerating the process. In stark contrast, for case (iii) the drift velocity acts towards $x_a$ [marked by the black arrow above curve (iii)] and thereby assists the first-passage; resetting at $x_0<\theta$ thus interrupts the process and delays it. Case (ii) represents an intermediate scenario between these two extreme cases, shown by curve (ii) in \fref{Fig4}. \\
\begin{figure}[b!]
\begin{centering}
\includegraphics[width=8.0cm]{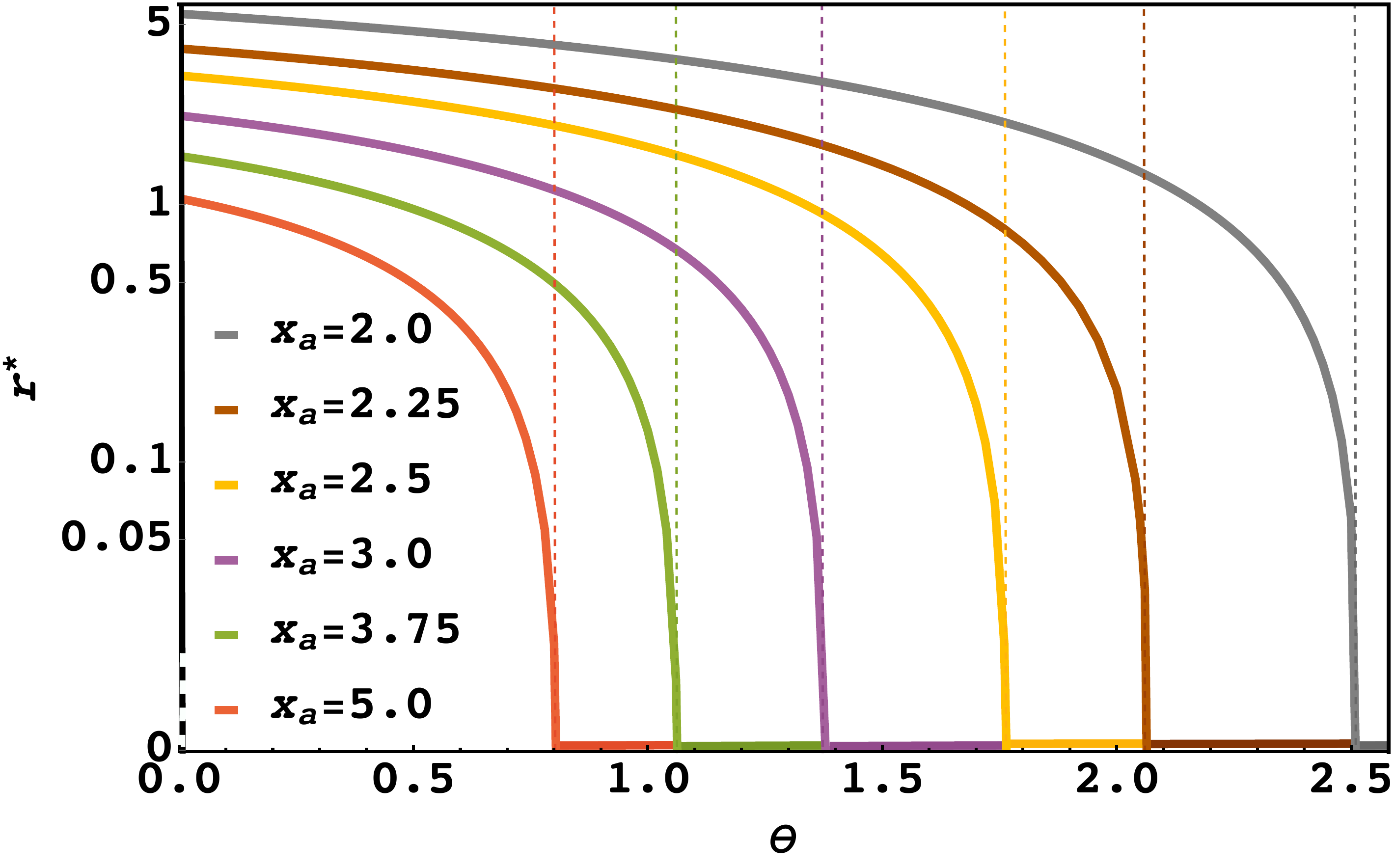}
\end{centering}
\caption{The optimal resetting rate $r^{\star}$ vs. $\theta$, obtained from the numerical solution of \eref{Trans_R}, for different positions of the absorbing boundary, $x_a$, when $x_0<x_a$. The non-zero values of $r^{\star}$, observed for lower values of $\theta$, mark the regime when resetting accelerates first-passage from $x_0$ to $x_a$. The resetting transitions are indicated by the dashed vertical lines with the same color as the curves. The initial position is kept fixed at $x_0=1$ in each case. With increase in $x_a$, the transition is observed to take place at lower values of $\theta$.
}
\label{Fig5}
\end{figure}
\indent
To develop a deeper understanding of the transition observed in \fref{Fig3}(a) and 3(b), we numerically solve \eref{Trans_R} to calculate $r^{\star}$ as a function of $\theta$ for different values of $x_a$ (keeping $x_0$ constant), and plot the same in \fref{Fig5}. It is clear from \fref{Fig5} that when $x_0$ is kept constant, as the distance of the absorbing boundary from the origin ($x_a$) increases, the transition appears at lower values of $\theta_c$. The three independent parameters, viz. $x_0$, $x_a$ and $\theta_c$, make the dynamics quite complicated to analyse, nonetheless, we try to understand this trend qualitatively as follows. Recall that for Feller process, the diffusion is inhomogeneous in space; the movement of the particle close to the origin is almost deterministic and it gradually becomes more and more diffusive when the particle moves away from the origin. In addition, the Feller potential owns a minimum at $x=\theta$. When the distance between $x_0$ and $x_a$ is small 
and both are placed somewhat close to the \textit{trapping zone} created by $\theta$, resetting at $x_0$ can reduce the otherwise long time spent in that trapping zone and accelerate the first-passage by incorporating some directed motion towards $x_a$. Keeping $x_0$ and $\theta$ unaltered, if the absorbing boundary is moved away from the origin (which is equivalent to moving vertically downward at any certain $\theta$ in \fref{Fig5}), the role of diffusion in the dynamics becomes more and more prominent. Thus, when the distance between $x_0$ and $x_a$ is large, diffusion near the absorbing boundary is quite high. Resetting the particle to a position $x_0$, where the dynamics is much less diffusive, clearly interrupts the first-passage as each resetting event makes the particle cross the trapping zone and travel the long distance between $x_0$ to $x_a$ all over again. This explains why the optimal resetting rate gradually decreases with increase in $x_a$ for any particular value of $\theta$ in \fref{Fig5}. The trapping becomes more significant as $\theta$ increases [as the potential well becomes deeper, see \fref{Fig1}] and that enhances the interruption that resetting triggers in case of longer travel distances. As a result, the resetting transition appears at lower $\theta_c$ as $x_a$ is placed further away from the origin. In fact, when $x_a$ is placed at a distance far enough, resetting at $x_0$ will only help if $\theta<x_0$, by successfully counteracting the trapping events at $x<x_0$. Summarizing, we see that for $x_0<x_a$, resetting expedites first-passage for $\theta<\theta_c$ and $\theta_c$ decreases as the distance to travel increases. After analysing \fref{Fig5} in a qualitative manner, we proceed to calculate the maximal speedup to quantify the effect of optimal resetting on the dynamics. \\
\subsection{Maximal speedup for $x_0<x_a$}
\begin{figure}[t!]
\begin{centering}
\includegraphics[width=8.0cm]{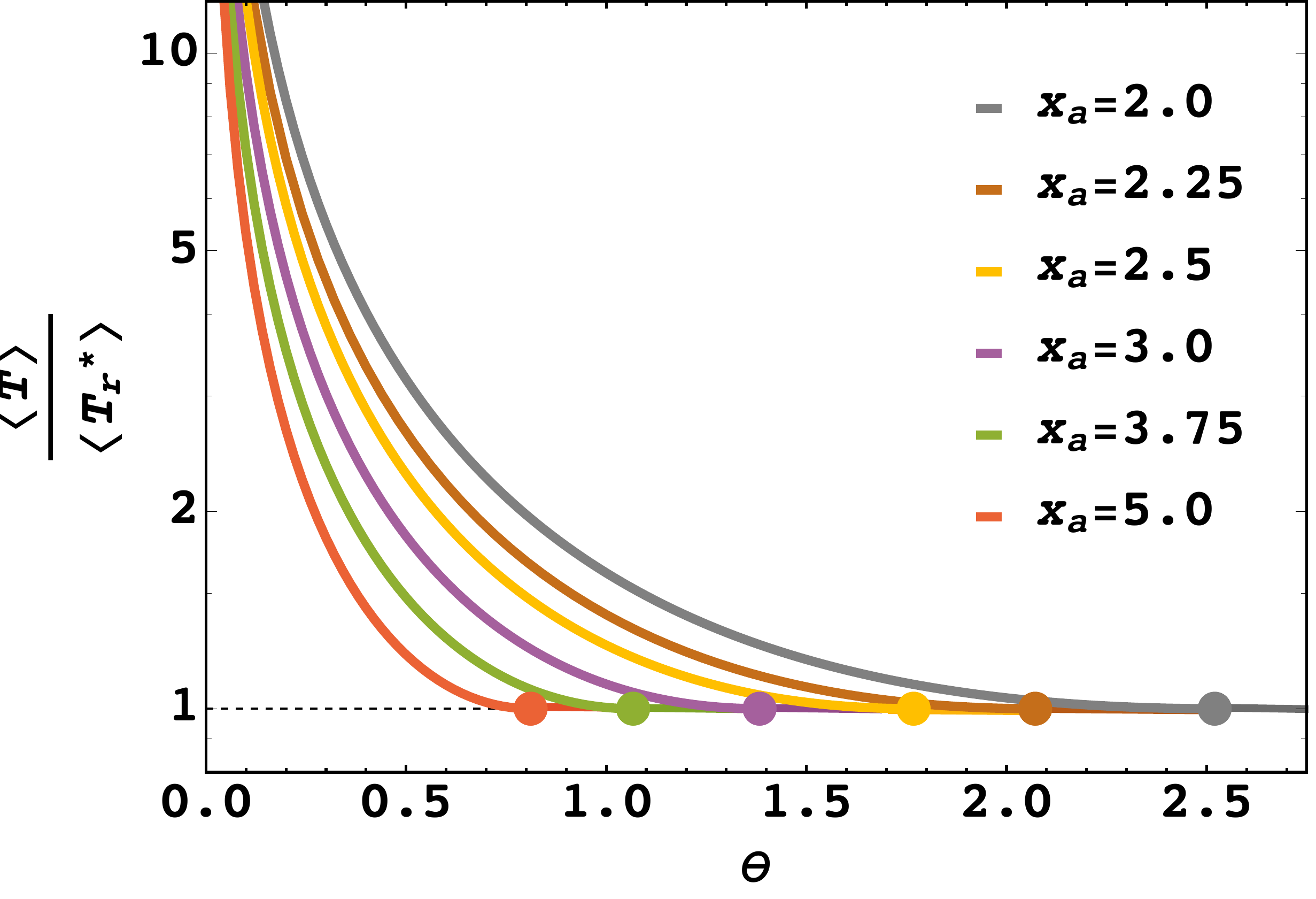}
\end{centering}
\caption{Maximal speedup $\left<T\right>/\left<T_{r^{\star}}\right>$ [due to resetting with an optimal rate $r^{\star}$] vs. $\theta$ following \eref{MSR} for different values of $x_a$, where $x_0<x_a$. The colors of the curves correspond to the cases shown in \fref{Fig5}. The most significant maximal speedup is observed for lower values of $\theta$. With gradual increase of $\theta$ past a critical value, the resetting transition [marked by colored discs] takes place and the maximal speedup becomes unity thereafter, which indicates that resetting no longer accelerate first-passage. } 
\label{Fig6}
\end{figure}
Resetting with an optimal rate renders the maximal speedup of a first-passage process. We define maximal speedup as the ratio between the mean FPT for the original process (i.e., the process without resetting) to the mean FPT of the process with optimal resetting, i.e, $\left<T\right>/\left<T_{r^{\star}}\right>$. Setting $r=r^{\star}$ in \eref{MFPT_R} and utilizing \eref{TR} from Appendix A [that gives the mean FPT of the original process], we can write down the following expression for the maximal speedup
\begin{align}
\frac{\left<T\right>}{\left<T_{r^{\star}}\right>}=\frac{r^{\star}\left(\left[\frac{\p M(s;\theta;x_a)}{\p s}\right]_{s\to 0}\mbox{\hspace{-0.45cm}} - \left[\frac{\p M(s;\theta;x_0)}{\p s}\right]_{s\to 0}\right)}{\frac{M(r^{\star};\theta;x_a)}{M(r^{\star};\theta;x_0)}-1}.
\label{MSR}
\end{align}
Plugging in $r^{\star}$ [that we calculated earlier by numerically solving \eref{Trans_R}] into \eref{MSR}, we calculate the maximal speedup of the first-passage process from $x_0$ to $x_a$, when $x_0<x_a$. Plotting \eref{MSR} with respect to $\theta$ for different values of $x_a$ in \fref{Fig6}, we see that the maximal speedup is most significant when $\theta$ is small, indicating that resetting in this regime helps the most. With increase in $\theta$, however, the maximal speedup gradually decreases, until it becomes unity at the point of resetting transition, where $r^{\star}=0$, as expected. \fref{Fig6} clearly shows that introduction of resetting with an optimal rate can even make the original first-passage process almost ten times faster!\\
\begin{figure*}[ht!]
\begin{centering}
\includegraphics[width=7.2cm]{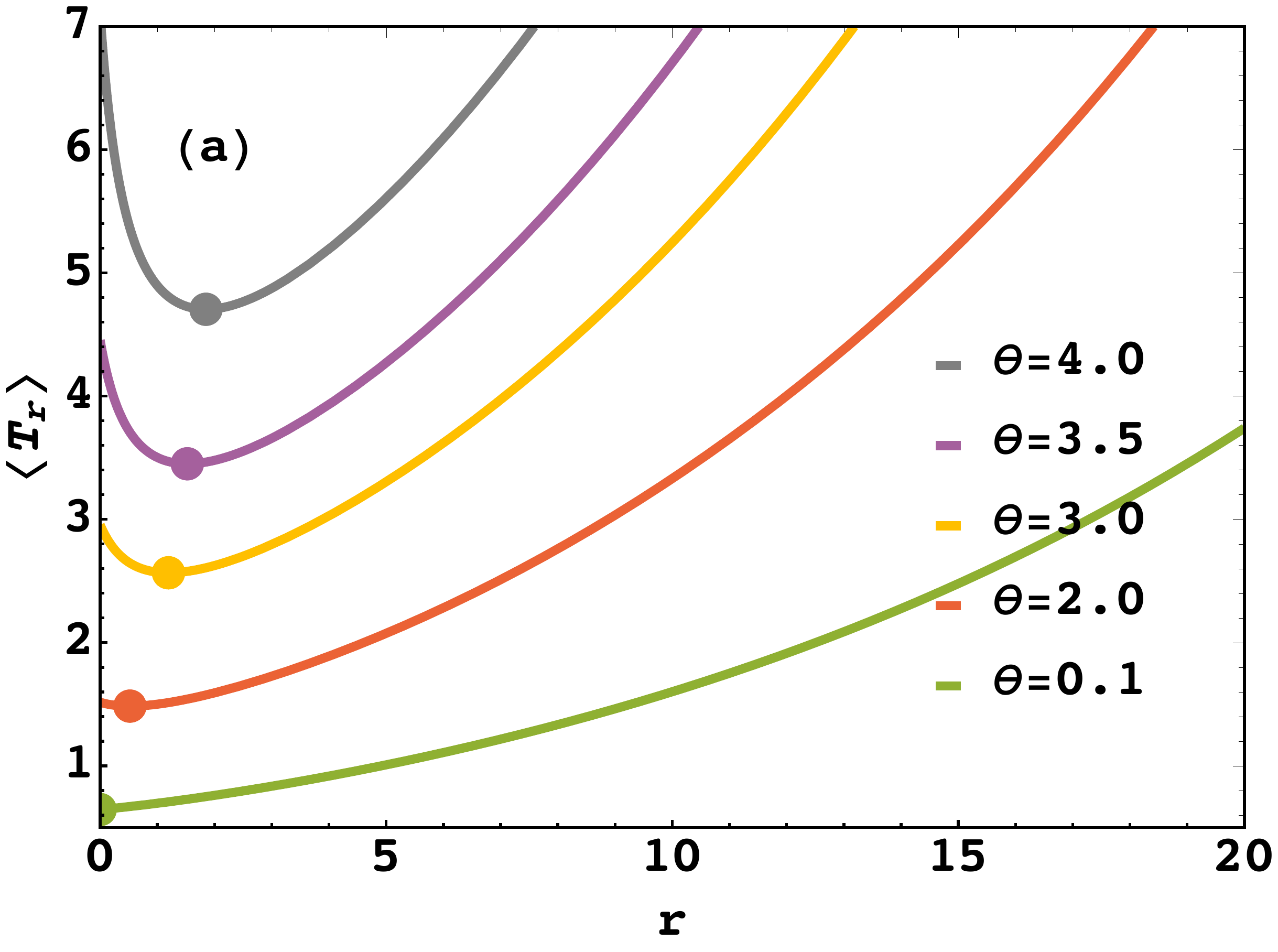}
\hspace{1.25cm}
\includegraphics[width=8.2cm]{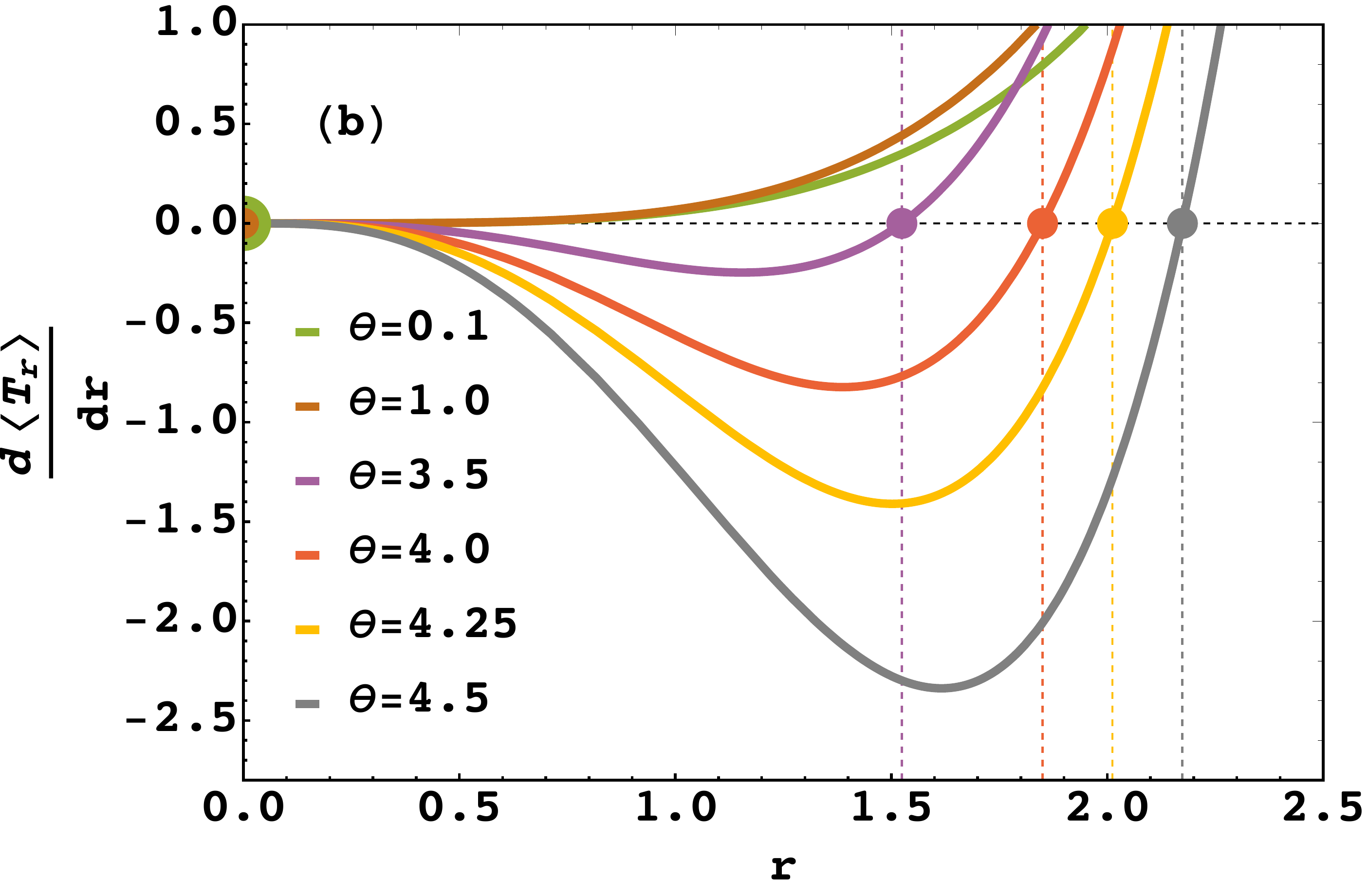}
\end{centering}
\caption{First-passage with resetting for $x_a<x_0$. Panel (a): The mean FPT $\left<T_r\right>$ vs. the resetting rate $r$ following \eref{MFPT_L} for different values of $\theta$, where $x_a=1$ and $x_0=2.5$. $\left<T_r\right>$ shows monotonic increase with $r$ for lower values of $\theta$, and the optimal resetting rates ($r^{\star}$), marked by colored discs, are zero in this regime. In contrast, for higher values of $\theta$, the variation of $\left<T_r\right>$ with $r$ becomes non-monotonic and $r^{\star}$ become non-zero.
Panel (b): Graphical solution of \eref{Trans_L} for different values of $\theta$, where $x_a=1$ and $x_0=2.5$. The solutions, denoted $r^{\star}$, are marked by colored discs. 
}
\label{Fig7}
\end{figure*}
\indent
Next, we proceed to explore the other scenario, viz., where the absorbing boundary ($x_a$) is placed to the left hand side of $x_0$.\\
\section{First-Passage from $x_0$ to $x_a$: when the boundary is placed closer to the origin compared to the initial position ($x_a<x_0$)}
\indent 
When $x_a<x_0$, the particle undergoes Feller diffusion in the interval $\Omega=[x_a,\infty)$. Note that the placement of the boundary closer to the origin compared to the initial position suggests that in this case, the first-passage is being considered from a more diffusive to a less diffusive zone. Moreover, in contrast to the previous case, studied in Section IV, now the domain $\Omega$ is \textit{semi-infinite}. Recalling the general expression of $Q_r(s|x_0)$ given in \eref{QrL_sol_g}, we see that in the limit $x_0\to \infty$, $M(s+r;\theta;x_0)$ diverges. Therefore, to keep $\tilde{Q}_r(s|x_0)$ finite for all values of $x_0$, we set $c_1=0$. The absorbing boundary at $x_a$ then gives $c_2=-\left[1+r\tilde{Q}_r(s|x_r)\right]/[(s+r)U(s+r;\theta;x_a)]$. Putting these values of $c_1$ and $c_2$ in \eref{QrL_sol_g}, we retrieve the specific solution for \eref{bfpe_qr} for $x_a<x_0$, which reads
\begin{align}
    \tilde{Q}_r(s|x_0)=\frac{1+r\tilde{Q}_r(s|x_r)}{s+r}\left[1-\frac{U(s+r;\theta;x_0)}{U(s+r;\theta;x_a)}\right].
    \label{Qr_solL}
\end{align}
In a similar manner as before, we set $x_r=x_0$ in \eref{Qr_solL}, i.e, coincide the position of resetting with the initial position, to obtain the following expression of $\tilde{Q}_r(s|x_0)$ in a self consistent manner
\begin{align}
    \tilde{Q}_r(s|x_0)=
    \frac{1-\frac{U(s+r;\theta;x_0)}{U(s+r;\theta;x_a)}}{s+r \left[\frac{U(s+r;\theta;x_0)}{U(s+r;\theta;x_a)}\right]}.
    \label{Qr_sol2L}
\end{align}

Note that in the absence of resetting, i.e, when $r\to 0$, \eref{Qr_sol2L} reduces to \eref{Q_solL}, as expected.
Setting $s=0$ in \eref{Qr_sol2L}, we obtain the mean first-passage time
\begin{equation}
    \left<T_r\right>=\frac{1}{r}\left[\frac{U(r,\theta,x_a)}{U(r,\theta,x_0)}-1\right].
    \label{MFPT_L}
\end{equation}\\ 
\indent
In \fref{Fig7}(a), we plot the mean FPT $\left<T_r\right>$ as a function of the resetting rate, $r$, following \eref{MFPT_L}. \fref{Fig7}(a) shows that [in complete contrast with the previous case, where $x_0<x_a$] for lower values of $\theta$, $\left<T_r\right>$ increases monotonically with $r$, which indicates that resetting can not expedite first-passage in that case. However, for higher values of $\theta$, $\left<T_r\right>$ show a non-monotonic variation, where the initial reduction of the mean FPT with the resetting rate indicates that resetting can successfully lower $\left<T_r\right>$ here. Therefore, the optimal resetting rate is zero for lower values of $\theta$ and that becomes non-zero 
when $\theta$ increases beyond a threshold value, $\theta_c$ [not shown in \fref{Fig7}(a)]. Next, we explore the resulting resetting transition in terms of the optimal resetting rate, i.e, the resetting rate that minimizes the mean FPT. \\
\subsection{The optimal resetting rate for $x_a<x_0$}
Letting $r^{\star}$ denote the optimal resetting rate as before, we differentiate \eref{MFPT_L} with $r$ and equate it to zero for $r=r^{\star}$ to obtain
\begin{widetext}
\begin{eqnarray}
\left[\frac{d \left<T_r\right>}{dr}\right]_{r=r^{\star}}\mbox{\hspace{-0.5cm}}=
\frac{1}{(r^{\star})^2 U(r^{\star};\theta;x_0)}
\left[
U(r^{\star};\theta;x_0)\left(U(r^{\star};\theta;x_0)
-U(r^{\star};\theta;x_a)
+r^{\star}\left[\frac{\p U(r;\theta;x_a)}{\p r}\right]_{r=r^{\star}}\right)\right.\nonumber\\
\left.-r^{\star}U(r^{\star};\theta;x_a)
\left[\frac{\p U(r;\theta;x_0)}{\p r}\right]_{r=r^{\star}}\right]=0.
    \label{Trans_L}
\end{eqnarray}
\end{widetext}
\eref{Trans_L} is a transcendental equation [like \eref{Trans_R}], and hence we need to solve it numerically in order to calculate $r^{\star}$. In \fref{Fig7}(b), we plot $\left[\frac{d \left<T_r\right>}{dr}\right]_{r=r^{\star}}$ vs. $r$ from \eref{Trans_L} to graphically solve the same in a similar way as in Sec. IV. 
\begin{figure}[ht!]
\includegraphics[width=7.2cm]{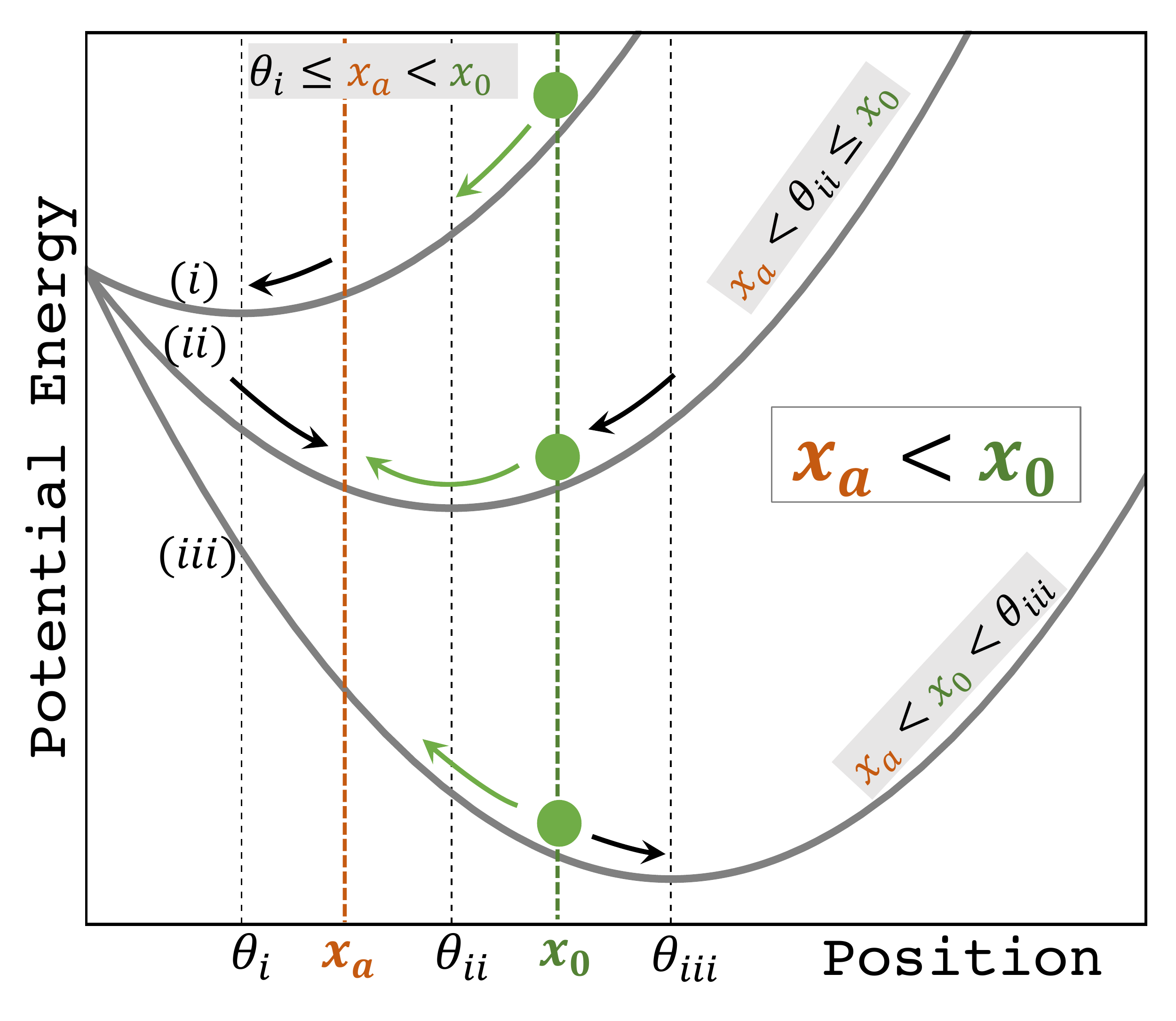}
\caption{Schematic plot of the potential $U(x)=x\left(\frac{x}{2}-\theta\right)$ vs. $x$ to mark the three scenarios that arise in connection to the first-passage from $x_0$ to $x_a$  for $x_a<x_0$, shown by three $U(x)$ curves with different minima, such that $\theta_i<\theta_{ii}<\theta_{iii}$. The green arrow in each case indicate the overall direction of first-passage, whereas the black arrows indicate the drift velocity generated by $U(x)$. 
When these forces assist each other, resetting is expected to delay first-passage [e.g., $\theta_i\le x_a<x_0$]. In contrast, when these forces oppose each other, resetting is expected to accelerate first-passage [e.g., $x_a<x_0<\theta_{iii}$].
An intermediate case is observed for $x_a<\theta_{ii}\le x_0$.
}
\label{Fig8}
\end{figure}
It is clear from \fref{Fig7}(b) that the optimal resetting rates are zero for $\theta\le\theta_c$, and resetting expedites the dynamics only when $\theta>\theta_c$, indicated by the non-zero values of $r^{\star}$. We can qualitatively explain this trend by considering Feller process as space-dependent diffusion in a unique harmonic potential $U(x)=x\left(\frac{x}{2}-\theta\right)$, as discussed earlier. For the present case, i.e, $x_a<x_0$, the relative placements of $\theta$ with $x_0$ and $x_a$ can create three distinct possibilities, viz., (i) $\theta\le x_a<x_0$, (ii) $x_a<\theta\le x_0$ and (iii) $x_a<x_0<\theta$. We illustrate these three cases in \fref{Fig8}, where we plot $U(x)$ vs. $x$ for three different values of $\theta$. Since $\theta$ represents the equilibrium position of $U(x)$, for case (i) the first-passage is a journey downhill [marked by the green arrow above curve (i)], while it is an uphill one for case (iii) [marked by the green arrow above curve (iii)]. As shown in \fref{Fig8}, the drift velocity acts towards the absorbing boundary for case (i) [marked by the black arrow above curve (i)], and thereby assists the first-passage; resetting at $x_0>\theta$ thus interrupts the original process and delays it. In contrast, the drift velocity for case (iii) acts away from $x_a$ [marked by the black arrow above curve (iii)], and thereby opposes the first-passage; resetting at $x_0<\theta$ thus accelerates the process. Case (ii) represents an intermediate scenario between these two extreme cases, shown by curve (ii) in \fref{Fig8}. \\
\begin{figure}[b!]
\begin{centering}
\includegraphics[width=8.0cm]{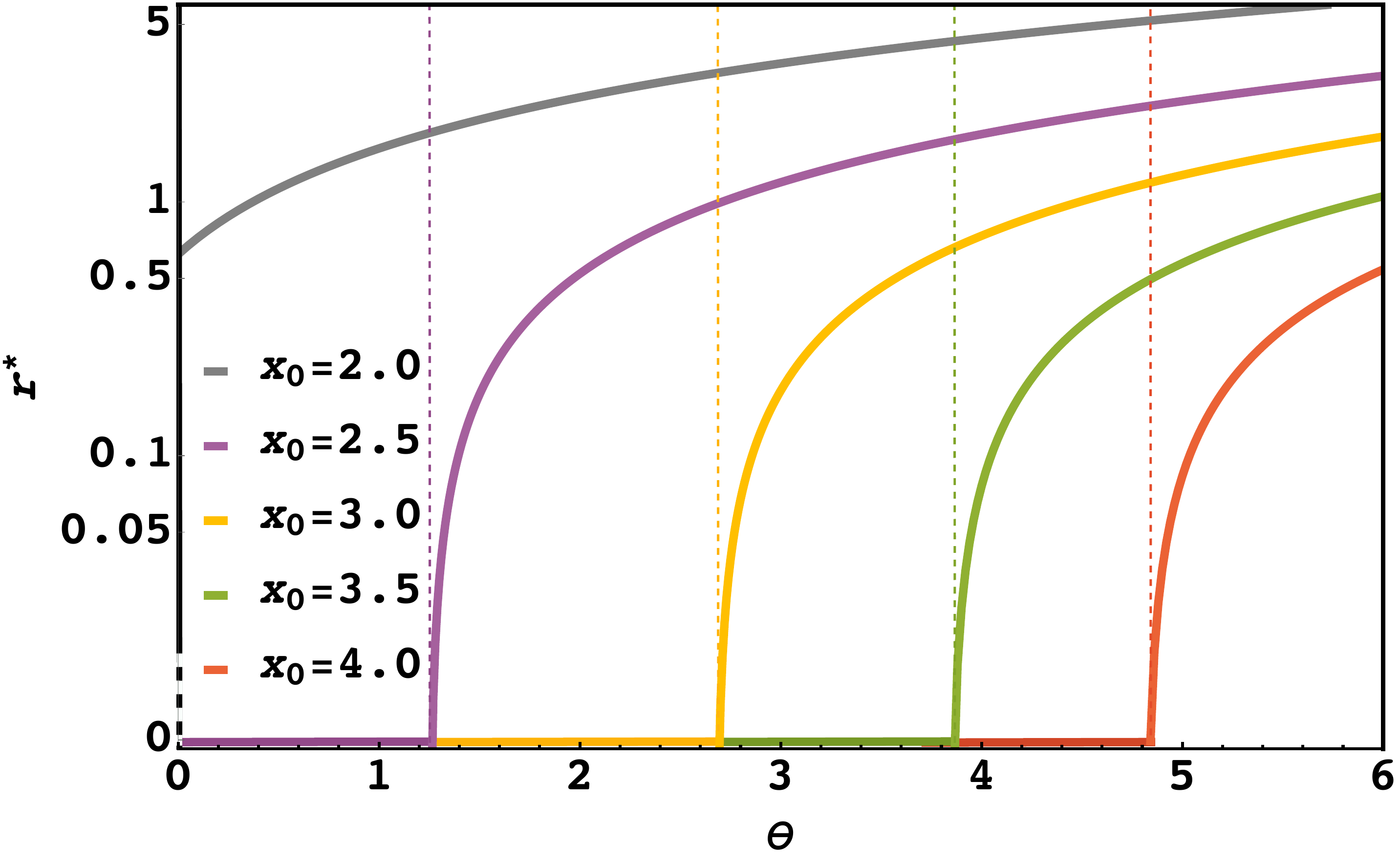}
\end{centering}
\caption{The optimal resetting rate $r^{\star}$, obtained from the numerical solution of \eref{Trans_L}, vs. $\theta$ for different initial positions, $x_0$, when $x_a<x_0$. The non-zero values of $r^{\star}$ mark the regime when resetting accelerates first-passage from $x_0$ to $x_a$, observed for higher values of $\theta$. The resetting transitions are clearly indicated by the dashed vertical lines of the same color as the curves. The position of the absorbing boundary is kept fixed at $x_a=1$ in each case. With increase in $x_0$, the transition is observed to take place at higher values of $\theta$. When the distance to travel is short enough, resetting always expedites first-passage, as observed for $x_0=2.0$ (gray curve).
}
\label{Fig9}
\end{figure}
\indent
To explore the resetting transition further, next we numerically solve \eref{Trans_L} to calculate $r^{\star}$ as a function of $\theta$ for different values of $x_0$, keeping $x_a$ constant. Plotting the results in \fref{Fig9}, we observe that when $x_0$ is considerably small, i.e., when the distance to travel is short, resetting  always expedites first-passage, which means resetting transition can even be non-existent! For higher values of $x_0$, however, the transition is observed and the threshold value of $\theta$ corresponding to the transition, i.e, $\theta_c$, increases with $x_0$. To qualitatively explain this trend, we recall that Feller process can be envisioned as \textit{space-dependent} diffusion [in a potential $U(x)=x(\frac{x}{2}-\theta)$] with a diffusion coefficient $D(x)=x$. This makes the potential \textit{effectively} asymmetric (or `tilted') around the equilibrium position $\theta$, such that the left branch of the potential appears a lot steeper to the particle compared to the right branch. Indeed, the particle diffuses almost freely when $x>>\theta$. Additionally, it can experience trapping around $\theta$, but that does not affect the dynamics as significantly as in the previous case for the following reasons: (a) for smaller values of $\theta$, the potential is shallow and hence not very effective in trapping the particle successfully and (b) for larger values of $\theta$, though the depth of the potential increases, the trapping is counterbalanced by the higher diffusion coefficient. Now, if the distance between the absorbing boundary and initial position is small and both are placed somewhat close to the origin, resetting at $x_0$ accelerates first-passage by cutting short the long trajectories that may generate as the particle diffuses far away from the origin. In contrast, when the distance between $x_a$ and $x_0$ is large [and $x_a$ is still kept close to the origin], resetting the particle in an almost free-diffusing zone far away from the boundary does not help expediting first-passage anymore unless $\theta$ is sufficiently high [see \fref{Fig8}]. Therefore, when $x_0$ is increased keeping $x_a$ unaltered, $\theta_c$ is observed to increase. Summarizing, we see that for $x_a<x_0$, resetting expedites first-passage for $\theta>\theta_c$ and $\theta_c$ increases as the distance to travel increases. Finally, we quantify the effect of optimal resetting on the dynamics, by calculating the maximal speedup, as we did earlier.
\subsection{Maximal speedup for $x_a<x_0$}
\begin{figure}[ht!]
\includegraphics[width=7.8cm]{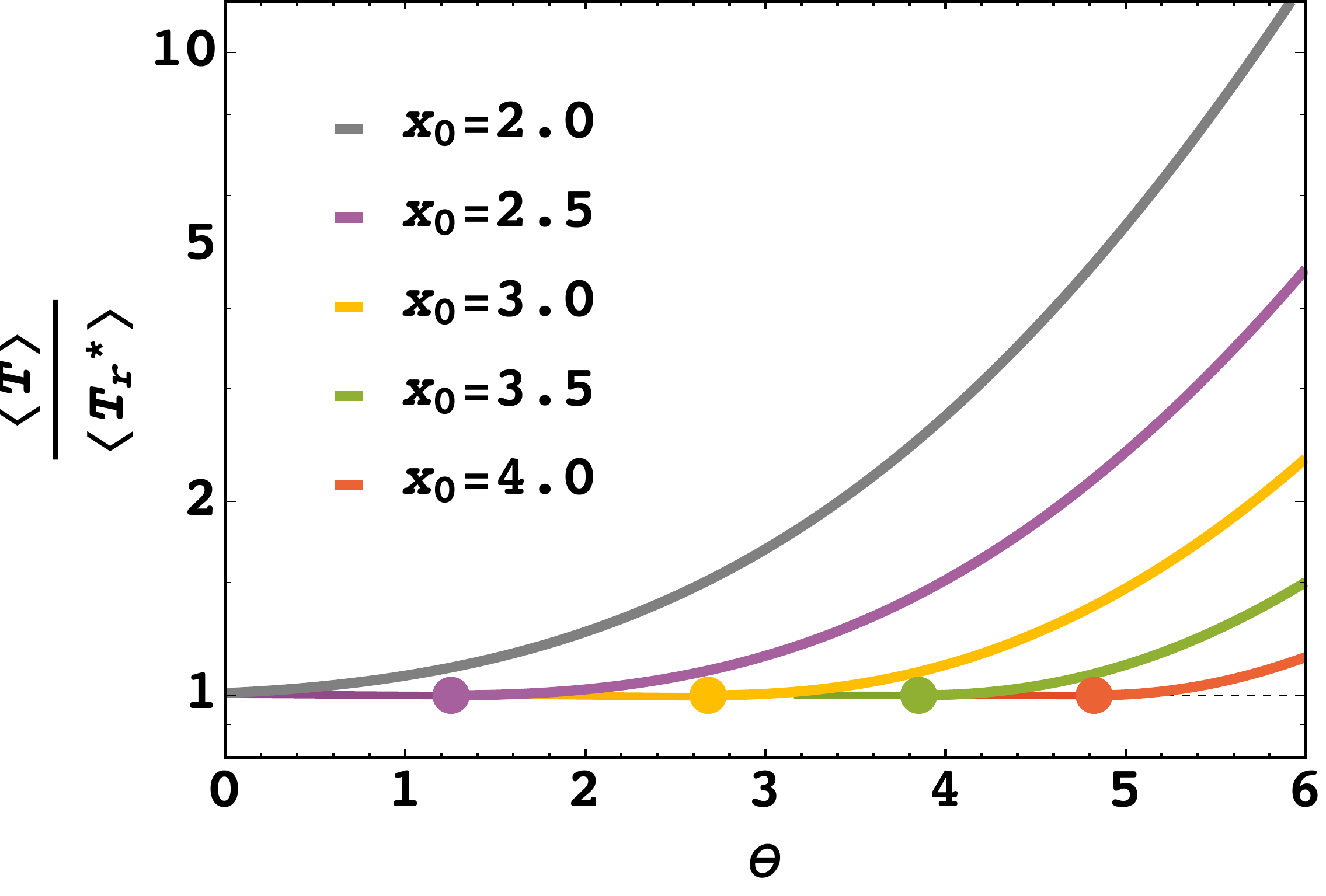}
\caption{Maximal speedup $\left<T\right>/\left<T_{r^{\star}}\right>$  vs. $\theta$ following \eref{MSL} for different values of $x_0$, where $x_a<x_0$. The colors of the curves correspond to the cases shown in \fref{Fig9}. In most cases (expect for the case of $x_0=2$, shown by the gray curve), maximal speedup is unity for lower values of $\theta$, which indicates that resetting does not expedite first-passage there. Most significant maximal speedup is observed for higher values of $\theta$. The resulting resetting transition are marked by colored discs. }
\label{Fig10}
\end{figure}
Since the maximal speedup of a first-passage process is defined as the ratio between the mean FPT for the original process (i.e., the process without resetting) to the mean FPT of the process with optimal resetting, setting $r=r^{\star}$ in \eref{MFPT_L} and using \eref{TL} from Appendix A, we obtain
\begin{align}
\frac{\left<T\right>}{\left<T_{r^{\star}}\right>}=\frac{r^{\star}\left(\left[\frac{\p U(s;\theta;x_a)}{\p s}\right]_{s\to 0}\mbox{\hspace{-0.45cm}} - \left[\frac{\p U(s;\theta;x_0)}{\p s}\right]_{s\to 0}\right)}{\frac{U(r^{\star};\theta;x_a)}{U(r^{\star};\theta;x_0)}-1}.
\label{MSL}
\end{align}
Plugging in $r^{\star}$ [obtained earlier by numerically solving \eref{Trans_L}] into \eref{MSL}, we calculate the maximal speedup of the first-passage process from $x_0$ to $x_a$, when $x_a<x_0$. Plotting \eref{MSL} with respect to $\theta$ for different values of $x_0$ in \fref{Fig10}, we see that the maximal speedup is most significant when $\theta$ is large, indicating that resetting in this regime helps the most. With decrease in $\theta$, however, the maximal speedup gradually decreases, until it becomes unity at the point of resetting transition, where $r^{\star}$ becomes zero. Note that for $x_0=2$ (shown by the grey curve in \fref{Fig9}), the maximal speedup is \textit{not} unity for small values of $\theta$; it just appears so in comparison to the significantly larger values of the maximal speedup that are observed for the higher values of $\theta$.
\section{Conclusions}
In this work, we explored the first-passage properties associated to Feller diffusion with Poissonian resetting. Considering Feller process as space-dependent diffusion with a diffusion coefficient $D(x)=x$ in a potential $U(x)=x\left(\frac{x}{2}-\theta\right)$, we calculated the first-passage time using the Fokker–Planck description of the system. Closed-form formul\ae \;were obtained for the Laplace transform of the survival probability, which in turn generated the exact expression of the mean FPT. We calculated the optimal resetting rates (rate of resetting that minimizes the mean FPT), which manifest a hallmark of {\it resetting transition} depending upon the governing parameters, viz., the equilibrium position of the Feller potential ($\theta$), the initial position of the particle executing Feller diffusion ($x_0$), and the position of the absorbing boundary that ensures the completion of the process ($x_a$). In-depth analysis were performed on the optimal resetting rate, $r^{\star}$, and the maximal speedup, $\left<T\right>/\left<T_{r^{\star}}\right>$, to identify the parameter space where Poissonian resetting accelerates Feller diffusion. The entire study was executed for two distinct cases: (a) when the absorbing boundary is placed further away from the origin compared to the initial position (or the starting value of the process lies below the target value, i.e., $x_0<x_a$) and (b) when the absorbing boundary is placed closer to the origin compared to the initial position (or the starting value of the process lies above the target value, i.e., $x_a<x_0$). For the former case, resetting was found to expedite first-passage from $x_0$ to $x_a$ for $\theta<\theta_c$, where $\theta_c$ is a critical value of $\theta$, which decreases when $x_a$ is moved away from the origin (and hence from $x_0$). In complete contrast, for the latter case resetting was observed to expedite first-passage for $\theta>\theta_c$, where $\theta_c$ is again a threshold value of $\theta$, which increases when $x_0$ is moved away from the origin (and hence from $x_a$). Interestingly enough, our study indicates that irrespective of that placement of the target value ($x_a$) either above or below the initial value ($x_0$) of the Feller process, the volume of the phase space where resetting expedites first-passage is always smaller when the distance to travel is large, which is evident from \fref{Fig4} and \fref{Fig9}. Since Feller process with resetting finds direct applications in various fields ranging from population dynamics to financial markets, we hope that the present work will attract attention from multiple disciplines associated to biological and social sciences.
\section*{Acknowledgements}
The author acknowledges the INSPIRE Faculty fellowship and research grant (IFA19-CH326) from the Department of Science \& Technology, Govt. of India, executed at IIT Tirupati through Project No. CHY/2021/005/DSTX/SOMR. Sincere thanks are due to MPIPKS, Dresden, Germany for hospitality during Summer, 2022. \\
\section*{Appendix A: Derivation of $CV(T)$, the coefficient of variation of first-passage time for Feller diffusion without resetting}
\renewcommand{\theequation}{A.\arabic{equation}}
\setcounter{equation}{0}
The coefficient of variation of the FPT is defined as the ratio of the standard deviation in the first-passage time $T$ to its mean, i.e., $CV(T)\coloneqq \sigma(T)/\left<T\right>$, where $\sigma(T)\coloneqq [\left<T^2\right>-\left<T\right>^2]^{\frac{1}{2}}$ is the standard deviation in $T$. Therefore, to calculate $CV(T)$, we need to calculate $\left<T\right>$ and $\left<T^2\right>$, i.e, the first and second moment of the FPT distribution. These two observables can be calculated from the survival probability of the 
associated first-passage process. Here we derive  approximate expressions of $CV(T)$ for the two different boundary conditions that are discussed in the main text, starting with $x_0<x_a$.\\ 
\indent
The mean first-passage time is related to the survival probability in the Laplace space as $\left<T\right>=\left[\tilde{Q}(s|x_0)\right]_{s\to 0}$. Expanding $M(s;\theta;x_i)$ [defined in \eref{M_def}] in Taylor series around $s=0$, 
for small values of $s$ we can write
\begin{eqnarray}
M(s;\theta;x_i)\sim 1+\left[\frac{\p M(s;\theta;x_i)}{\p s}\right]_{s\to 0}\mbox{\hspace{-0.45cm}}s+\frac{1}{2}\left[\frac{\p^2 M(s;\theta;x_i)}{\p s^2}\right]_{s\to 0}\mbox{\hspace{-0.45cm}}s^2,\nonumber\\
\label{M_Taylor}
\end{eqnarray}
where $i\equiv 0,a$. Plugging in that into \eref{Q_solR} we get 
\onecolumngrid
\noindent
 \begin{align}
    \tilde{Q}(s|x_0)\sim
    \frac{1}{s}\left[1-\frac{1+\left[\frac{\p M(s;\theta;x_0)}{\p s}\right]_{s\to 0}\mbox{\hspace{-0.25cm}}s+\frac{1}{2}\left[\frac{\p^2 M(s;\theta;x_0)}{\p s^2}\right]_{s\to 0}\mbox{\hspace{-0.25cm}}s^2}{1+\left[\frac{\p M(s;\theta;x_a)}{\p s}\right]_{s\to 0}\mbox{\hspace{-0.25cm}}s+\frac{1}{2}\left[\frac{\p^2 M(s;\theta;x_a)}{\p s^2}\right]_{s\to 0}\mbox{\hspace{-0.25cm}}s^2}\right].
    \label{QR_Taylor}
\end{align}  
In the limit $s\to 0$, \eref{QR_Taylor} gives an approximate expression for the mean FPT of Feller diffusion for $x_0<x_a$, which reads
\begin{align}
\left<T\right>=\left[\frac{\p M(s;\theta;x_a)}{\p s}\right]_{s\to 0}\mbox{\hspace{-0.45cm}} - \left[\frac{\p M(s;\theta;x_0)}{\p s}\right]_{s\to 0}.
\label{TR}
\end{align}
Since $\left<T^2\right>=-2\left[\frac{d\tilde{Q}(s|x_0)}{ds}\right]_{s\to0}$, differentiating \eref{QR_Taylor} with respect to $s$ and setting the limit $s\to 0$, we obtain
\begin{align}
\left<T^2\right>=
&2
\left[\left[\frac{\p M(s;\theta;x_a)}{\p s}\right]_{s\to 0}\mbox{\hspace{-0.4cm}}
-\left[\frac{\p M(s;\theta;x_0)}{\p s}\right]_{s\to 0}\right]\mbox{\hspace{-0.2cm}}\left[\frac{\p M(s;\theta;x_a)}{\p s}\right]_{s\to 0}\nonumber\\
&+\left[\frac{\p^2 M(s;\theta;x_0)}{\p s^2}\right]_{s\to 0}\mbox{\hspace{-0.45cm}} - \left[\frac{\p^2 M(s;\theta;x_a)}{\p s^2}\right]_{s\to 0}.
\mbox{\hspace{-0.4cm}}
\label{Tsq_R}
\end{align}
Utilizing \eref{TR} and \eref{Tsq_R}, we obtain an expression for $CV(T)$ that reads
\onecolumngrid
\begin{align}
CV(T)= \frac{\sqrt{\left[\frac{\p^2 M(s;\theta;x_0)}{\p s^2}\right]_{s\to 0}\mbox{\hspace{-0.45cm}} - \left[\frac{\p^2 M(s;\theta;x_a)}{\p s^2}\right]_{s\to 0}+
\left(\left[\frac{\p M(s;\theta;x_a)}{\p s}\right]_{s\to 0}\right)^2
- \left(\left[\frac{\p M(s;\theta;x_0)}{\p s}\right]_{s\to 0}\right)^2} }
{\left[\frac{\p M(s;\theta;x_a)}{\p s}\right]_{s\to 0}\mbox{\hspace{-0.45cm}} - \left[\frac{\p M(s;\theta;x_0)}{\p s}\right]_{s\to 0}}\;\;\;\;\;\mbox{for}\;\;\;\;\;x_0<x_a.
\label{CVR}
\end{align}
We calculate $CV(T)$ following \eref{CVR} and plot that in \fref{Fig2}(a) in the main text. \\
\indent
In a similar manner, for $x_a<x_0$, we can expand $U(s;\theta;x_i)$ [defined in \eref{U_def}] in Taylor series around $s=0$ and for small values of $s$ that reads
\begin{align}
U(s;\theta;x_i)\sim 1+\left[\frac{\p U(s;\theta;x_i)}{\p s}\right]_{s\to 0}\mbox{\hspace{-0.45cm}}s+\frac{1}{2}\left[\frac{\p^2 U(s;\theta;x_i)}{\p s^2}\right]_{s\to 0}\mbox{\hspace{-0.45cm}}s^2,
\label{U_Taylor}
\end{align}
for $i\equiv 0,a$. Plugging in the expression of $U(s;\theta;x_i)$ from \eref{U_Taylor} into \eref{Q_solL}, we obtain
 \begin{eqnarray}
    \tilde{Q}(s|x_0)\sim
    \frac{1}{s}\left[1-\frac{1+\left[\frac{\p U(s;\theta;x_0)}{\p s}\right]_{s\to 0}\mbox{\hspace{-0.25cm}}s+\frac{1}{2}\left[\frac{\p^2 U(s;\theta;x_0)}{\p s^2}\right]_{s\to 0}\mbox{\hspace{-0.25cm}}s^2}{1+\left[\frac{\p U(s;\theta;x_a)}{\p s}\right]_{s\to 0}\mbox{\hspace{-0.25cm}}s+\frac{1}{2}\left[\frac{\p^2 U(s;\theta;x_a)}{\p s^2}\right]_{s\to 0}\mbox{\hspace{-0.25cm}}s^2}\right],
    \label{QL_Taylor}
\end{eqnarray}  
which the limit of $s\to 0$ gives
\begin{align}
\left<T\right>= \left[\frac{\p U(s;\theta;x_a)}{\p s}\right]_{s\to 0}\mbox{\hspace{-0.45cm}} - \left[\frac{\p U(s;\theta;x_0)}{\p s}\right]_{s\to 0}.
\label{TL}
\end{align}
Differentiating \eref{QL_Taylor} with respect to $s$, in the limit $s\to 0$ we obtain
\begin{align}
\left<T^2\right>=
&2
\left[\left[\frac{\p U(s;\theta;x_a)}{\p s}\right]_{s\to 0}\mbox{\hspace{-0.5cm}}
-\left[\frac{\p U(s;\theta;x_0)}{\p s}\right]_{s\to 0}\right]
\mbox{\hspace{-0.2cm}}
\left[\frac{\p U(s;\theta;x_a)}{\p s}\right]_{s\to 0}\nonumber\\
&+\left[\frac{\p^2 U(s;\theta;x_0)}{\p s^2}\right]_{s\to 0}\mbox{\hspace{-0.45cm}} - \left[\frac{\p^2 U(s;\theta;x_a)}{\p s^2}\right]_{s\to 0}.
\mbox{\hspace{-0.4cm}}
\label{Tsq_L}
\end{align}
From \eref{TL} and \eref{Tsq_L}, we finally get \vspace{0.2cm}

\begin{align}
CV(T)= \frac{\sqrt{\left[\frac{\p^2 U(s;\theta;x_0)}{\p s^2}\right]_{s\to 0}\mbox{\hspace{-0.45cm}} - \left[\frac{\p^2 U(s;\theta;x_a)}{\p s^2}\right]_{s\to 0}+
\left(\left[\frac{\p U(s;\theta;x_a)}{\p s}\right]_{s\to 0}\right)^2
- \left(\left[\frac{\p U(s;\theta;x_0)}{\p s}\right]_{s\to 0}\right)^2} }
{\left[\frac{\p U(s;\theta;x_a)}{\p s}\right]_{s\to 0}\mbox{\hspace{-0.45cm}} - \left[\frac{\p U(s;\theta;x_0)}{\p s}\right]_{s\to 0}}\;\;\;\;\;\mbox{for}\;\;\;\;\;x_a<x_0.
\label{CVL}
\end{align}

We calculate $CV(T)$ for $x_a<x_0$ from \eref{CVL} and plot the same in \fref{Fig2}(b) in the main text.
\vspace{0.5cm}
\twocolumngrid
 
\end{document}